\documentclass[preprint,groupedaddress,pre]{revtex4}
\usepackage{amssymb}
\usepackage{amsmath}

\input{tcilatex}

\begin{document}

\title{Stiff three-frequency orbit of the hydrogen atom }
\author{Jayme De Luca}
\email[author's email address:]{ deluca@df.ufscar.br}
\affiliation{Universidade Federal de S\~{a}o Carlos, \\
Departamento de F\'{\i}sica\\
Rodovia Washington Luis, km 235\\
Caixa Postal 676, S\~{a}o Carlos, S\~{a}o Paulo 13565-905}
\date{\today }

\begin{abstract}
We study a stiff quasi-periodic orbit of the electromagnetic two-body
problem of Dirac's electrodynamics of \emph{point }charges. The delay
equations of motion are expanded about circular orbits to obtain the
variational equations up to nonlinear terms. The three-frequency orbit
involves two harmonic modes of the variational dynamics with a period of the
order of the time for light to travel the interparticle distance. In the
atomic magnitude, these harmonic modes have a frequency that is fast
compared with the circular rotation. The quasi-periodic orbit has three
frequencies; the frequency of circular rotation (\emph{slow}) and the two
fast frequencies of two mutually orthogonal harmonic modes. Poynting's
theorem gives a mechanism for a beat of the mutually orthogonal fast modes
to cancel the radiation of the unperturbed circular motion by interference.
The nonradiation condition for destructive interference is that the fast
frequencies beat precisely at the circular frequency. The resonant orbits
have magnitudes in qualitative and quantitative agreement with quantum
electrodynamics (QED), as follows; (i) the orbital angular momenta are
integer multiples of Planck's constant to a good approximation, (ii) the
orbital frequencies agree with a corresponding emission line of QED within a
few percent on average, (iii) the orbital frequencies are given by a
difference of two linear eigenvalues, viz., the frequencies of the mutually
orthogonal fast modes, and (iv) the angular momentum of gyration of the
variational motion about each resonant orbit is of the order of Planck's
constant.
\end{abstract}

\maketitle

\section{Introduction}

We study a stiff quasi-circular orbit of the electromagnetic two-body
problem of Dirac's electrodynamics with retarded-only fields\cite{Dirac}, a
dynamics with implicitly-defined delay. The motivation is to understand this
complex dynamics described by the delay equations for particle separations
in the atomic magnitude\cite{dissipaFokker}. We give an economical method to
derive the variational equations of dynamics about circular orbits up to
nonlinear terms. The particular harmonic solutions of the variational
dynamics, with a period of the order of the time for light to travel the
interparticle distance, are henceforth called the \emph{ping-pong} modes
(PP). For atomic orbits, the frequency of the PP modes turns out to be much
faster than the orbital frequency. \ These modes introduce a fast (\emph{%
stiff}) timescale in the dynamics, and are physically important for the
particles to have the option to avoid radiating energy away. The PP modes
for vibration along the orbital plane turn out to have \emph{almost} the
same frequency of PP modes for vibrations normal to the orbital plane, a
remarkable quasi-degeneracy that naturally produces beats with a frequency
of the order of the orbital frequency. The quasi-periodic orbits of hydrogen
have three frequencies; the slow frequency of the circular orbit plus the
two fast frequencies corresponding to a planar PP mode and a perpendicular
PP mode, as illustrated in Fig. 1. This special combination appears in a
mechanism suggested by Poynting's theorem to cancel the radiation of the
slow circular motion by interference with a beat oscillation of the two
mutually orthogonal PP modes. We investigate the conditions for the
nonlinear variational equations to accept such fast gyrating solutions about
the circular orbit, i.e., the PP oscillations. After the fast dynamics is
established, Poynting's theorem gives a necessary resonance condition to
avoid radiative losses; viz., the two mutually orthogonal PP modes must beat
at the orbital frequency. This resonance condition turns out to be satisfied
precisely in the atomic magnitude. The fast gyration defines an angular
momentum vector of the order of the orbital angular momentum of the
unperturbed circular orbit. We stress that the point charges are not
spinning about themselves, but rather gyrating about a guiding-center that
is moving along a slow circular orbit, as illustrated in Fig. 1. The stiff
three-frequency orbits share several magnitudes with those of the hydrogen
atom of quantum electrodynamics (QED) \cite{Bohr}, with reasonable precision
and qualitative detail. The circular frequency of a resonant orbit agrees
with the corresponding line of QED within a few percent average deviation.
There is also a large body of qualitative agreement with QED; (i) the
resonant orbits have orbital angular momenta that are approximate integer
multiples of a basic angular momentum. This basic angular momentum agrees
well with Planck's constant, (ii) the angular momentum of gyration of the
variational motion about the circular orbit is of the order of Planck's
constant. This angular momentum of gyration is a vector that rotates at the
orbital frequency, and\ (iii) the emitted frequency is given by a difference
of two linear eigenvalues, i.e., the frequencies of the PP modes,
analogously to the Rydberg-Ritz principle of QED.

The equations of motion of Dirac's electrodynamics of \emph{point} charges%
\cite{Dirac} are briefly discussed in Appendix A. After Dirac's 1938 work%
\cite{Dirac}, an early study of Eliezer \cite{Eliezer, Parrott, Andrea,
Massimo} revealed the surprising result that an electron moving in an
attractive Coulomb field can never fall into the center of force by
radiating energy (henceforth called Eliezer's theorem). It was subsequently
found that only scattering states are possible in any tridimensional motion
with self-interaction in a Coulomb field \cite{Andrea, Massimo}. Eliezer's
theorem strongly suggests that a finite mass for the proton is essential for
a physically meaningful dynamics in the electromagnetic two-body problem.
When the proton has an infinite mass, there is an inertial frame where it
rests at all times, and in this frame the protonic field on the electron is
simply a Coulomb field, i.e., the dynamics of Eliezer's theorem. On the
other hand, if the proton has a \emph{finite} mass, such inertial frame does
not exist and the equations of motion involve delay, because of the finite
speed of light. A finite mass for the proton is what brings delay into the
electromagnetic two-body dynamics, with its associated ping-pong phenomenon.
The infinite-mass limit is a singular limit, because the equations of motion
pass from delay equations to ordinary differential equations. This work is
an attempt to put together what is lost in this singular limit where the PP
modes disappear. We stress that the circular orbit is not an exact solution
of the full equations of motion, so that we are not doing Lyapunov
stability, but rather constructing particular solutions of the nonlinear
variational equations, i.e., the three-frequency orbit.

The road map for this paper is as follows. In Appendix A we review the
electrodynamics of point charges in a generalized setting that includes
Dirac's theory as a special case, and give the equations of motion of point
charges in an intuitive form. A non-specialist reader should start reading
the paper from Appendix A. In Section II we define the PP modes and the
quantities of the circular orbit, to be used as an approximate solution in
Sections III, IV and V. In Section III we outline our economical method to
derive the variational equations, a method that expands the implicit
light-cone condition and uses the action formalism. In Section IV we derive
the linearized variational equations and study the PP modes of tangent
dynamics for vibrations along the orbital plane. This derivation is
laborious and makes full use of our economical method plus the use of a
symbolic manipulations software. We derive the linearized variational
equations in the generalized electromagnetic setting of Appendix A, to
compare with previously known results, but we stress that in Section V we
use only Dirac's electrodynamics, the physically interesting special case.
In Section V we give an application of Dirac's electrodynamics to atomic
physics, by discussing the existence of a three-frequency orbit involving
mutually orthogonal PP oscillations of finite amplitude, i.e., a particular
solution of the variational equations. We investigate the mechanism to
cancel the radiation of the circular orbit by interference with a beat of
two mutually orthogonal PP modes, a mechanism that starts to operate
immediately after the PP dynamics is established. We discuss the necessary
resonance condition of Poynting's theorem to avoid radiation, i.e., that the
PP modes beat at the orbital frequency. We study the resonant orbits that
are stabilized by this mechanism and compare their magnitudes with the
magnitudes of QED. In Section V we also give a second derivation of
Poynting's resonance condition. This derivation averages the angular
momentum of gyration over the fast timescale, yielding a vector rotating at
a slow frequency. The rotating angular momentum produces a gyroscopic torque
on the slow dynamics, and introduces the same resonance condition of
Poynting's theorem. In Appendix B we derive the tangent dynamics for
oscillations perpendicular to the orbital plane, analogously to what is done
in Section IV for the planar variational equations. These two derivations
can be given separately up to the linear order, but the $z$ and $xy$
variational equations are otherwise coupled at higher order. The existence
of fast harmonic solutions of finite amplitude is discussed in Appendix C.
In Appendix C we also give a third derivation of the resonance condition, a
derivation based on the detailed balance of the guiding-center dynamics.
Last, in Section VI we put the conclusions and discussion.

\bigskip

\section{The circular orbit}

\bigskip

Our perturbation scheme takes the circular orbit as a first approximation.
We use the index $i=1$ to indicate quantities of the electron and $i=2$ for
the proton, with masses $m_{1}$ and $m_{2}$, respectively. We henceforth use
a unit system where the speed of light is $c\equiv 1$, the electronic charge
is $e_{1}=-e_{2}\equiv -1$ and the mass of the electron is $m_{1}\equiv 1$.
In our unit system, the mass of the proton is given by $m_{2}=1824$,
approximately. The circular orbit is illustrated in Fig. 2; the particles
move in concentric circles and in diametral opposition at the same time of
the inertial frame. The details of \ the familiar Coulombian circular orbit
will be given now. The constant angular velocity is indicated by $\Omega $,
the interparticle distance in light-cone is $r_{b}$ and the angle that one
particle turns while the light emanating from the other particle reaches it
is $\theta \equiv \Omega r_{b}$. The delay angle $\theta $ is the natural
independent parameter of the circular orbit, which turns out to be small for
orbits of the atomic magnitude, $\theta \lesssim 10^{-2}$. For small $\theta 
$, the interparticle distance in light-cone, $r_{b}$, is $O(\theta ^{2})$
close to the interparticle distance at the same time of the inertial frame, $%
r_{0}$. Because of this $O(\theta ^{2})$ approximation, the familiar
Coulombian formulas with $r_{b}$ replaced \ by $r_{0}$ yield the leading
order formulas in powers of $\theta $. For example, the orbital frequency,
Kepler's law, is given to leading-order in $\theta $ by 
\begin{equation}
\Omega =\mu \theta ^{3}+\cdots ,  \label{Kepler}
\end{equation}%
where $\mu \equiv m_{1}m_{2}/(m_{1}+m_{2})$ is the reduced mass (here and
henceforth). In our unit system, $\mu =(1824/1825)\simeq 1$ for hydrogen.
The interparticle distance in light-cone, $r_{b}$, is constant along the
circular orbit and given, to leading order, by 
\begin{equation}
r_{b}=\frac{1}{\mu \theta ^{2}}+\cdots .  \label{RB}
\end{equation}%
Using the radial equation of motion for the Coulombian orbit, one finds that
the angular momentum of the circular orbit is given, to leading order in $%
\theta $, by%
\begin{equation}
l_{z}=\frac{1}{\theta }+\cdots .  \label{angular-momentum}
\end{equation}%
The units of $l_{z}$ in Eq. (\ref{angular-momentum}) are $e^{2}/c$, just
that we are using a unit system where $e^{2}=c=1$. For orbits of the atomic
magnitude, $l_{z}$ as defined by Eq. (\ref{angular-momentum}) is of the
order of one over the fine-structure constant, $\alpha ^{-1}=137.036$. Each
particle travels a circular orbit with radius and scalar velocity defined by 
\begin{eqnarray}
r_{1} &\equiv &b_{1}r_{b},  \label{defradius} \\
r_{2} &\equiv &b_{2}r_{b},  \notag
\end{eqnarray}%
and 
\begin{eqnarray}
v_{1} &=&\Omega r_{1}=\theta b_{1},  \label{defvelocity} \\
v_{2} &=&\Omega r_{2}=\theta b_{2},  \notag
\end{eqnarray}%
for the electron and for the proton, respectively. For consistency with the
definition of $r_{b}$ as the exact interparticle distance in light-cone, the
definition of $b_{1}$ and $b_{2}$ must include an $O(\theta ^{2})$ term; $%
b_{1}\equiv (1+g\theta ^{2})m_{2}/(m_{1}+m_{2})$ and $b_{2\text{ }}\equiv
(1+g\theta ^{2})m_{1}/(m_{1}+m_{2})$. Evaluating the interparticle distance
in light-cone and equating it to $r_{b}$ yields 
\begin{equation}
r_{b}^{2}=r_{b}^{2}(b_{1}^{2}+b_{2}^{2}+2b_{1}b_{2}\cos (\theta )).
\label{circularcone}
\end{equation}%
Formula (\ref{circularcone}) becomes Eq. (3.1) of Ref.\cite{Schild} after
use of Eq. (\ref{defvelocity}). Expanding Eq. (\ref{circularcone}) for small 
$\theta $ yields $g\simeq 0.5\mu /(m_{1}+m_{2})$. The ratio $%
(b_{1}/b_{2})=(m_{2}/m_{1})$ is the Coulombian ratio and the radii defined
by Eq.(\ref{defradius}) are $O(\theta ^{2})$ near the Coulombian radii, and
we henceforth call this the Coulombian circular orbit. As discussed in
Appendix C, \ for Dirac's electrodynamics the circular orbit is only an
approximate solution. Since we are doing perturbation, it suffices to use
the above defined Coulombian orbit as an approximate solution, and let the
perturbation scheme take care of the correction. In the action-at-a-distance
theory\cite{Schild,Hans}, it turns out that the circular orbit \emph{is} an
exact orbit of the two-body problem. The action-at-distance electrodynamics
is a special setting described by the equations of Appendix A with $\Gamma
=-1/2$, and is used here only to crosscheck the method of Section III.

The intuitive picture of a PP oscillation is a ping-pong game, i.e., the
particles throwing a ball back and forth at the finite speed $c=1$, as a
means to communicate changes in position. The PP modes have a period of the
order of $r_{b}/c$, where $r_{b}$ is given by Eq. (\ref{RB}). The order of
magnitude of the PP frequency is

\begin{equation}
w_{PP}\sim \frac{2\pi }{r_{b}/c}=2\pi \mu \theta ^{2},  \label{WPP}
\end{equation}%
where we used that $c=1$ in our unit system. To compare the PP frequency
with the Coulombian orbital frequency (\ref{Kepler}), it is useful to
express the PP frequency (\ref{WPP}) as

\begin{equation}
w_{PP}\sim 2\pi \mu \theta ^{2}\equiv \frac{|\lambda |}{\theta }\Omega ,
\label{defstiff}
\end{equation}%
where $|\lambda |$ is a number of the order of $2\pi $. For atomic orbits, $%
\theta \simeq 10^{-2}$, the PP frequency $w_{PP}$ is larger than $\Omega $
by three orders of magnitude. We henceforth define the generalized frequency
of a ping-pong normal mode by $\lambda \Omega /\theta $, so that a purely
imaginary $\lambda $ defines a harmonic oscillation, and the eventual real
part of $\lambda $ defines a damping or a runaway. For small $\theta $, the
bouncing-time for light to travel back and forth the interparticle distance
is approximately $2r_{b}/c$, so that the phase-shift of the fast oscillation
during the bouncing-time is $\Delta \Phi \equiv \func{Im}(\frac{\lambda
\Omega }{\theta }\frac{r_{b}}{c})$. This phase-shift evaluates to $\Delta
\Phi =2\func{Im}(\lambda )$ by use of $c=1$ and $\Omega r_{b}=\theta $, and
we shall see in Section V and in Appendix C that $\lambda $ must be an
integer multiple of $2\pi $, i.e., $\lambda =\pi q$ . The integer number $q$
is henceforth called the ping-pong index.

\section{\protect\bigskip Variational expansion about a circular orbit}

\ 

\ The variational equations are obtained by substituting a circular orbit
plus a perturbation into the equations of motion of Appendix A, e.g., Eq. (%
\ref{Eqfamiliar}), and expanding in powers of the perturbation. Since these
are complex equations, even after using a convenient coordinate system this
derivation is long. Our method uses the fact that the equations of motion
are \emph{almost} the Euler-Lagrange equations of a suitable Lagrangian,
with the addition of the self-interaction force. After the variational
equations of the Lagrangian sector are derived, we add in the variation of
the terms due to the self-interaction. Readers should consult Appendix A for
a brief review of the equations of motion of point charges and definition of
terms such as the Lorentz-force, the Lorentz-Dirac self-force and the action
integral. In this Section we outline the procedure of substituting a
circular orbit plus a perturbation into action (\ref{VAintegr}) of Appendix
A and expanding the action up to a desired order. Minimization of this
truncated action plus the expansion of the kinetic action yields the Lorentz
sector of the variational equations. Since we perform the algebraic
computations with a symbolic manipulations software, it is equally easy to
derive the variational equations in a generalized electromagnetic setting
that contains Dirac's electrodynamics as a special case, as explained in
Appendix A. The calculation in the generalized setting has some interest of
its own, and provides a useful cross-check with other known calculations 
\cite{dissipaFokker,astar2B}. The generalized setting of Appendix A has an
arbitrary constant $\Gamma $ in the Green's function, and we stress that $%
\Gamma $ will be set to\emph{\ zero } in our application of Section V, i.e.,
we study the hydrogen orbit in Dirac's electrodynamics with retarded-only
fields \cite{Dirac}, a fundamental physical theory. A reader not interested
in this generality can assume $\Gamma =0$ throughout the whole paper.

The variational equations for planar perturbations of circular orbits
decuple from the equation for transverse perturbations up to the linear
order. It is convenient to write these planar variational equations using
complex gyroscopic coordinates rotating at the frequency $\Omega $ of the
unperturbed circular orbit. The coordinates $(x_{k},y_{k})$ of each particle
are defined by two complex gyroscopic coordinates $\eta _{k}$ and $\xi _{k}$
as 
\begin{eqnarray}
u_{j} &\equiv &x_{k}+iy_{k}\equiv r_{b}\exp (i\Omega t)[d_{k}+2\eta _{k}],
\label{coordinates} \\
u_{j}^{\ast } &\equiv &x_{k}-iy_{k}\equiv r_{b}\exp (-i\Omega t)[d_{k}+2\xi
_{k}],  \notag
\end{eqnarray}%
where $k=1$ for electron and $k=2$ for proton and the real quantities $%
d_{1}\equiv b_{1}$ and $d_{2}\equiv -b_{2}$ are defined in Eqs. (\ref%
{defradius}). Because $x_{k}$ and $y_{k}$ are real, we must have $\eta _{k}$ 
$\equiv \xi _{k}^{\ast }$, but to obtain the variational equations it
suffices to treat $\eta _{k}$ and $\xi _{k}$ as independent variables in
Lagrangian (\ref{VAC}). Two quantities appear so often in the calculations
that it is useful to name them; (i) The numerator of Lagrangian (\ref{VAC}),
evaluated along the circular orbit, is constant and given by 
\begin{equation}
C\equiv 1+b_{1}b_{2}\theta ^{2}\cos (\theta ),  \label{defC}
\end{equation}%
having the same value for both retarded and advanced interactions, and\ (ii)
The denominator of Lagrangian (\ref{VAC}), evaluated along the circular
orbit and divided by $r_{b}$, is constant and defined by

\begin{equation}
S\equiv 1+b_{1}b_{2}\theta \sin (\theta ),  \label{defS}
\end{equation}%
having also the same value for both retarded and advanced interactions. For
\ the stiff limit of Sections IV and V we shall ignore the $O(\theta ^{2})$
corrections and set $C$ $=S=1$. Here we derive the electron's equations of
motion only ($j=1$ in Eq. (\ref{coordinates})). The equation for the proton
is completely symmetric and can be obtained by interchanging indices $1$ and 
$2$. Notice that the exchange operation on the $d^{\prime }s$ is $\
d_{1}\Longleftrightarrow -d_{2}$. This is because $d_{1}=b_{1}$ is defined
positive while $d_{2}=-b_{2}$ is defined negative, so that at the same time
the particles are in diametral opposition along the unperturbed circular
orbit, as illustrated in Fig. 2.

The velocity of the electron at its time $t_{1}$ is calculated with Eq. (\ref%
{coordinates}) as

\begin{eqnarray}
\dot{u}_{1} &\equiv &v_{1x}+iv_{1y}\equiv \theta \exp (i\Omega
t_{1})[id_{1}+2i(\eta _{1}-i\dot{\eta}_{1}],  \label{velocity1} \\
\dot{u}_{1}^{\ast } &\equiv &v_{1x}-iv_{1y}\equiv \theta \exp (-i\Omega
t_{1})[-id_{1}-2i(\xi _{1}+i\dot{\xi}_{1})].  \notag
\end{eqnarray}%
\ The velocity of the proton at time $t_{2}$ can be obtained by
interchanging indices $1$ and $2$ in Eq. (\ref{velocity1}), as explained
above. Using the $c=\pm 1$ convention explained above Eq. (\ref{lightcone})
of Appendix A, the quantities of particle $2$ appearing in Lagrangian (\ref%
{VAC}) are evaluated at a time $t_{2}=t_{2c}$, in light-cone with the
position of particle $1$ at time $t_{1}$. The implicit light-cone condition
must be expanded and solved by iteration, and for that it is convenient to
define in each case the excess-lag function $\varphi _{c}$ by 
\begin{equation}
t_{2c}\equiv t_{1}+\frac{r_{b}}{c}+\frac{\varphi _{c}}{\Omega }.
\label{light-cone}
\end{equation}%
If the perturbation is zero, then $\varphi _{+}=\varphi _{-}=0$ and the
dynamics is along the original circular orbit, where the light-cone lag is
the constant $r_{b}$ for the advanced case and $-r_{b}$ for the retarded
case. We henceforth measure the evolution with the scaled-time parameter $%
\tau \equiv \Omega t_{1}$. The implicit definition of $\varphi _{c}$ by the
light-cone condition involves the position of particle $2$ at either the
advanced time $t_{2+}$ or the retarded time $t_{2-}$ , as defined by Eq. (%
\ref{coordinates}),

\begin{eqnarray}
u_{2}(\tau +c\theta +\varphi _{c}) &\equiv &r_{b}\exp (i\Omega
t_{2c})[d_{2}+2\eta _{2}(\tau +c\theta +\varphi _{c})],  \label{particle2a}
\\
u_{2}^{\ast }(\tau +c\theta +\varphi _{c}) &\equiv &r_{b}\exp (-i\Omega
t_{2c})[d_{2}+2\xi _{2}(\tau +c\theta +\varphi _{c})],  \notag
\end{eqnarray}%
as well as the velocity of particle $2$ at the advanced/retarded position;

\begin{eqnarray}
\dot{u}_{2}(\tau +c\theta +\varphi _{c}) &\equiv &i\theta \exp (i\Omega
t_{2c})[d_{2}+(\eta _{2}-i\dot{\eta}_{2})+2\varphi _{c}(\dot{\eta}_{2}-\ddot{%
\eta}_{2})],  \label{velocity2a} \\
\dot{u}_{2}^{\ast }(\tau +c\theta +\varphi _{c}) &\equiv &-i\theta \exp
(-i\Omega t_{2c})[d_{2}+(\xi _{2}+i\dot{\xi}_{2})+2\varphi _{c}(\dot{\xi}%
_{2}+\ddot{\xi}_{2})].  \notag
\end{eqnarray}

To obtain the linearized variational equations we expand the equations of
motion to first order in $\eta _{k}$ and $\xi _{k}$, which are the
Euler-Lagrange equations of the quadratic expansion of action (\ref{VAintegr}%
) in $\eta _{k}$ and $\xi _{k}$. We must therefore carry all expansions up
to the second order in the $\eta \xi $ coordinates. For example, the
position of particle $2$ is determined up to the second order by expanding
the arguments of $\eta _{2}$ and $\xi _{2}$ of Eq. (\ref{particle2a}) in a
Taylor series about the unperturbed light-cone for one order only as%
\begin{eqnarray}
u_{2}(\tau +c\theta +\varphi _{c}) &\simeq &r_{b}\exp (i\Omega
t_{2c})\{d_{2}+2[\eta _{2}(\tau +c\theta )+\varphi _{c}\dot{\eta}_{2}(\tau
+c\theta )]\},  \label{expansionz2a} \\
u_{2}^{\ast }(\tau +c\theta +\varphi _{c}) &\simeq &r_{b}\exp (-i\Omega
t_{2c})\{d_{2}+2[\xi _{2}(\tau +c\theta )+\varphi _{c}\dot{\xi}_{2}(\tau
+c\theta )]\}.  \notag
\end{eqnarray}%
In the following we find that $\varphi _{c}$ is linear in $\eta $ and $\xi $
to leading order, so that the next term in expansion (\ref{expansionz2a})
would be a third order term, not needed for the linear variational
equations. We henceforth indicate $\xi \eta $ quantities of particle $2$
evaluated at the \emph{unperturbed} light-cone by a subindex $c$, i.e., $%
\eta _{2c}\equiv $ $\eta _{2}(\tau +c\theta )$ and $\xi _{2c}\equiv \xi
_{2}(\tau +c\theta )$. Otherwise when the subindex $c$ appears, it indicates
evaluation on the exact light-cone. Notice that the small parameter of
expansion is the size of $\eta $ and $\xi $, and we henceforth expand any
quantity evaluated on the light-cone in a Taylor series about the \emph{%
unperturbed} light-cone, up to the order needed. For quasi-circular orbits
this method yields variational equations with a fixed delay. We stress that
one should never expand the arguments in powers of $\theta $; such an
expansion yields ordinary differential equations and looses the PP modes!
The perturbed light-cone is expressed implicitly by the distance from the
advanced/retarded position of particle $2$ to the present position of
particle $1$, described in gyroscopic coordinates by the modulus of the
complex number%
\begin{equation}
u\equiv u_{1}(\tau )-u_{2}(\tau +c\theta +\varphi _{c}),  \label{deFU}
\end{equation}%
where $c=1$ describes the advanced light-cone and $c=-1$ describes the
retarded light-cone. Using Eq. (\ref{expansionz2a}) to calculate $u$ up to
the second order yields 
\begin{eqnarray}
u &=&r_{b}\exp (i\Omega t_{2c})\{D_{c}^{\ast }+2[\exp (-ic\theta -i\varphi
_{c})\eta _{1}(\tau )-\eta _{2c}-\varphi _{c}\dot{\eta}_{2c}]\},
\label{separation} \\
u^{\ast } &=&r_{b}\exp (-i\Omega t_{2c})\{D_{c}+2[\exp (ic\theta +i\varphi
_{c})\xi _{1}(\tau )-\xi _{2c}-\varphi _{c}\dot{\xi}_{2c}]\},  \notag
\end{eqnarray}%
where we defined the following complex function of $\varphi _{c}$ 
\begin{equation}
D_{c}\equiv b_{2}+b_{1}\exp (ic\theta +i\varphi _{c}).  \label{defD}
\end{equation}%
At $\varphi _{c}=0$ (the unperturbed circular orbit), $D_{c}$ has a unitary
modulus, expressing the unperturbed light-cone condition (\ref{circularcone}%
). The right-hand side of Eq. (\ref{separation}) is a quadratic form times $%
\exp (i\Omega t_{2c})$, and in the action it appears multiplied by a
counter-rotating term, i.e., a quadratic form times $\exp (-i\Omega t_{2c})$%
, so that the product is independent of $t_{2c}$. Because of this rotational
symmetry of action (\ref{VAintegr}), the following quadratic Gauge
simplification can be applied directly to any quadratic rotating form. One
can integrate by parts a quadratic term of the quantity, e.g., equation (\ref%
{separation}), and disregard the boundary term. This Gauge simplification
yields a correct action up to the second order. For example, integrating by
parts the quadratic terms in $\dot{\eta}_{2c}$ and $\dot{\xi}_{2c}$ on the
right-hand side of Eq. (\ref{separation}) and disregarding the quadratic
Gauge yields%
\begin{eqnarray}
u &=&r_{b}\exp (i\Omega t_{2c})\{D_{c}^{\ast }+2[(1-i\varphi _{c})\exp
(-ic\theta )\eta _{1}(\tau )-(1-\dot{\varphi}_{c})\eta _{2c}]\},  \label{x}
\\
u^{\ast } &=&r_{b}\exp (-i\Omega t_{2c})\{D_{c}+2[(1+i\varphi _{c})\exp
(ic\theta )\xi _{1}(\tau )-(1-\dot{\varphi}_{c})\xi _{2c}]\},  \notag
\end{eqnarray}%
where we have also expanded $\exp (i\varphi _{c})$ up to the linear order in 
$\varphi _{c}$, enough to give the correct quadratic action. Analogously,
the velocity of particle $2$ has the following expansion up to a quadratic
Gauge%
\begin{eqnarray}
\dot{u}_{2}(\tau +c\theta +\varphi _{c}) &\simeq &i\theta \exp (i\Omega
t_{2c})[d_{2}+2(\eta _{2c}-i\dot{\eta}_{2c})(1-\dot{\varphi}_{c})],
\label{velocity2Gauged} \\
\dot{u}_{2}^{\ast }(\tau +c\theta +\varphi _{c}) &\simeq &-i\theta \exp
(-i\Omega t_{2c})[d_{2}+2(\xi _{2c}+i\dot{\xi}_{2c})(1-\dot{\varphi}_{c})]. 
\notag
\end{eqnarray}

Using the above quantities, the numerator of Lagrangian (\ref{VAC}) can be
calculated as%
\begin{equation}
h_{2}=(1-\mathbf{v}_{1}\cdot \mathbf{v}_{2c})=(1-\frac{1}{2}\dot{u}_{1}\dot{u%
}_{2}^{\ast }-\frac{1}{2}\dot{u}_{1}^{\ast }\dot{u}_{2}),  \label{h2}
\end{equation}%
and the denominator of Lagrangian (\ref{VAC}) can be calculated as%
\begin{equation}
h_{4}=r_{12c}(1+\frac{\mathbf{n}_{12c}\cdot \mathbf{v}_{2c}}{c}%
)=r_{b}(1+\phi )+\frac{u\dot{u}_{2}^{\ast }}{2c}+\frac{u^{\ast }\dot{u}_{2}}{%
2c}.  \label{h4}
\end{equation}%
In Eq. (\ref{h4}) we have introduced the scaled delay function $\phi $ by 
\begin{equation}
\varphi _{c}\equiv c\theta \phi ,  \label{scaledphi}
\end{equation}%
where the subindex under $\phi $ is omitted for simplicity of notation. To
relate $\phi $ to the $\xi \eta $ perturbations we expand the implicit
light-cone condition of the perturbed orbit up to the quadratic order%
\begin{equation}
uu^{\ast }=|x_{2}(t_{1}+\frac{r_{b}}{c}+\varphi
_{c})-x_{1}(t_{1})|^{2}=c^{2}(t_{2c}-t_{1})^{2}=(r_{b}+\frac{\varphi }{%
\Omega c})^{2}.  \label{quadra}
\end{equation}%
The light-cone condition (\ref{quadra}) is most simply expressed in terms of
the scaled $\phi $ defined in Eq. (\ref{scaledphi}). This expansion of Eq. (%
\ref{quadra}) up to the second order yields 
\begin{eqnarray}
C\phi ^{2}+2S\phi &=&2[(b_{1}\xi _{1}-b_{2}\xi _{2})+(b_{1}\eta
_{1}-b_{2}\eta _{2})]  \notag \\
&&+2[(b_{2}\xi _{1}-b_{1}\eta _{2})\exp (ic\theta )+(b_{2}\eta _{1}-b_{1}\xi
_{2})\exp (-ic\theta )]  \notag \\
&&+4[\xi _{1}\eta _{1}-\xi _{1}\eta _{2}\exp (ic\theta )-\xi _{2}\eta
_{1}\exp (-ic\theta )]  \notag \\
&&-2\varphi \lbrack b_{2}(\dot{\xi}_{2}+\dot{\eta}_{2})+b_{1}\dot{\xi}%
_{2}\exp (-ic\theta )+b_{1}\dot{\eta}_{2}\exp (ic\theta )].
\label{quadraticform}
\end{eqnarray}%
The solution to Eq. (\ref{quadraticform}) up to the first order in the $\xi
\eta $ coordinates is 
\begin{equation}
S\phi _{(1)}\equiv \lbrack (b_{1}\xi _{1}-b_{2}\xi _{2})+(b_{1}\eta
_{1}-b_{2}\eta _{2})+(b_{2}\xi _{1}-b_{1}\xi _{2})\exp (ic\theta
)+(b_{2}\eta _{1}-b_{1}\eta _{2})\exp (-ic\theta )].  \label{PHI1}
\end{equation}

Last, as a check for the above calculations, in the following we derive the
equations of motion for the circular orbit of the action-at-a-distance
theory \cite{Schild}. The Lagrangian of action (\ref{VAintegr}) for $\Gamma
=-1/2$ is $\pounds $ $\equiv (L_{+}+L_{-})/2$, and its expansion up to the
first order is 
\begin{equation}
\tilde{\pounds }=1-\{[\theta ^{2}(S-1)S+C^{2}](b_{1}+b_{2}\cos (\theta
))+S(\theta \sin (\theta )-\theta ^{2}\cos (\theta ))b_{2}\}\frac{(\eta
_{1}+\xi _{1})}{CS^{2}},  \label{linearVA}
\end{equation}%
where the tilde indicates that $\pounds $ was scaled by its value along the
unperturbed circular orbit, $\tilde{\pounds }\equiv r_{b}S$\ $\pounds /C$.
Scaling the kinetic Lagrangian with the same factor and expanding to first
order yields 
\begin{equation}
\tilde{T}_{1}=\frac{-r_{b}m_{1}S}{C\gamma _{1}}+\frac{r_{b}\theta
^{2}m_{1}\gamma _{1}Sb_{1}}{C}(\eta _{1}+\xi _{1}),  \label{linearkinetic}
\end{equation}%
where $\gamma _{1}\equiv 1/\sqrt{1-v_{1}^{2}\text{ }}$and $v_{1}$ is given
by Eq. (\ref{defvelocity}). The effective Lagrangian for particle $1$ up to
the linear order is%
\begin{equation}
\tilde{L}_{eff}^{(1)}\equiv \tilde{T}_{1}+\tilde{\pounds }.
\label{action1linear}
\end{equation}%
Lagrangian (\ref{action1linear}) is a linear functional of $\xi _{1}$,
independent of $\dot{\xi}_{1}$, so that the Euler-Lagrange equation for $\xi
_{1}$ is simply $\frac{\partial \tilde{L}_{eff}^{(1)}}{\partial \xi _{1}}=0$%
, i.e.,%
\begin{equation}
m_{1}b_{1}r_{b}\gamma _{1}\theta ^{2}S^{3}=[C^{2}+\theta
^{2}S(S-1)](b_{1}+b_{2}\cos (\theta ))+S(\theta \sin (\theta )-\theta
^{2}\cos (\theta ))b_{2}.  \label{Schild3.2}
\end{equation}%
This is Eq. 3.2 of Ref. \cite{Schild}, and the equation for $\eta _{1}$ is
the same condition by symmetry \ (this is actually the reason why the
circular orbit is a solution\cite{Schonberg}). The equation for particle $2$
is obtained by interchanging indices $1$ and $2$ in Eq. (\ref{Schild3.2}),
as is\ Eq. 3.3 of Ref. \cite{Schild}. In the next section we expand the
action to second order, to determine the linearized variational equations.

\section{\protect\bigskip Linearized variational equations}

In this Section we obtain the linear-order terms of the variational
equations. For this we carry the expansion of Section III to the quadratic
order. The next term of expansion (\ref{linearkinetic}) of the kinetic
Lagrangian of particle $1$, calculated with Eq. (\ref{velocity1}), is%
\begin{eqnarray}
\tilde{T}_{1} &=&T_{o}-\frac{r_{b}Sm_{1}}{C}\sqrt{1-|v_{1}|^{2}}=\frac{%
r_{b}\theta ^{2}m_{1}\gamma _{1}Sb_{1}}{C}(\eta _{1}+\xi _{1})
\label{kinetic1} \\
&&+\frac{Sr_{b}\theta ^{2}m_{1}\gamma _{1}}{2C}[\gamma _{1}^{2}(\xi _{1}+i%
\dot{\xi}_{1}+\eta _{1}-i\dot{\eta}_{1})^{2}-(\xi _{1}+i\dot{\xi}_{1}-\eta
_{1}+i\dot{\eta}_{1})^{2}]+\cdots .  \notag
\end{eqnarray}%
This kinetic Lagrangian of particle $1$ has a quadratic form defined by two
coefficients

\begin{eqnarray}
\tilde{T}_{1} &=&\frac{r_{b}\theta ^{2}m_{1}\gamma _{1}Sb_{1}}{C}(\eta
_{1}+\xi _{1})  \label{quadrat1} \\
&&+M_{1}[\dot{\eta}_{1}\dot{\xi}_{1}+i(\eta _{1}\dot{\xi}_{1}-\xi _{1}\dot{%
\eta}_{1})+\eta _{1}\xi _{1}]+\frac{\theta ^{2}G_{1}}{2}[\xi _{1}^{2}+\eta
_{1}^{2}-\dot{\xi}_{1}^{2}-\dot{\eta}_{1}^{2}],  \notag
\end{eqnarray}%
where $M_{1}\equiv (1+\gamma _{1}^{2})m_{1}\gamma _{1}r_{b}\theta ^{2}S/C$
and $G_{1}\equiv (\gamma _{1}^{2}-1)m_{1}\gamma _{1}\theta ^{2}S/C$. We also
need the solution of Eq. (\ref{quadraticform}) to second order, $\phi =\phi
_{(1)}+$ $\phi _{(2)}$ , where $\phi _{(1)}$ is given by Eq. (\ref{PHI1})
and $\phi _{(2)}$ is calculated by iteration to be

\begin{eqnarray}
S\phi _{(2)} &=&\varphi (ib_{2}\xi _{1}-ib_{1}\eta _{2}-b_{1}\dot{\eta}%
_{2})\exp (ic\theta )+\varphi (ib_{1}\xi _{2}-ib_{2}\eta _{1}-b_{1}\dot{\xi}%
_{2})\exp (-ic\theta )  \label{PHI2} \\
&&-\varphi b_{2}(\dot{\xi}_{2}+\dot{\eta}_{2})-\frac{1}{2}C[\phi _{(1)}]^{2}
\notag \\
&&+2[\xi _{1}\eta _{1}-\xi _{1}\eta _{2}\exp (ic\theta )-\xi _{2}\eta
_{1}\exp (-ic\theta )].  \notag
\end{eqnarray}%
Next we expand the numerator and the denominator of Lagrangian (\ref{VAC}),
Eqs. (\ref{h2}) and (\ref{h4}), up to the quadratic order. We also need the
following quantities expanded up to second order; (i) the particle
separation in light-cone, Eq. (\ref{separation}), (ii) the velocity of
particle $1$, Eq. (\ref{velocity1}), and (iii) the velocity of particle $2$
at time $t_{2c}$, Eq. (\ref{velocity2Gauged}). This quadratic form can be
greatly simplified by adding suitable Gauge terms to it (essentially
integration by parts), which yields the following \ normalized\emph{\ }%
Lagrangian for particle\emph{\ }$1$ 
\begin{eqnarray}
L_{c} &=&-\frac{i}{2}B_{21}(\eta _{1}\dot{\xi}_{1}-\xi _{1}\dot{\eta}%
_{1})+U_{11}\xi _{1}\eta _{1}+\frac{1}{2}N_{11}\xi _{1}^{2}+\frac{1}{2}%
N_{11}^{\ast }\eta _{1}^{2}  \label{quadraVA} \\
&&+R_{c}\xi _{1}\xi _{2c}+R_{c}^{\ast }\eta _{1}\eta _{2c}+P_{c}\xi _{1}\eta
_{2c}+P_{c}^{\ast }\eta _{1}\xi _{2c}+  \notag \\
&&+\frac{Y_{c}}{2}(\xi _{1}\dot{\xi}_{2c}-\xi _{2c}\dot{\xi}_{1})+\frac{%
Y_{c}^{\ast }}{2}(\eta _{1}\dot{\eta}_{2c}-\eta _{2c}\dot{\eta}_{1})+  \notag
\\
&&\frac{\Lambda _{c}}{2}(\xi _{1}\dot{\eta}_{2c}-\eta _{2c}\dot{\xi}_{1})+%
\frac{\Lambda _{c}^{\ast }}{2}(\eta _{1}\dot{\xi}_{1}-\xi _{2}\dot{\eta}%
_{1})+  \notag \\
&&+\frac{T_{c}}{2}\dot{\xi}_{1}\dot{\xi}_{2c}+\frac{T_{c}^{\ast }}{2}\dot{%
\eta}_{1}\dot{\eta}_{2c}+\frac{E_{c}}{2}\dot{\xi}_{1}\dot{\eta}_{2c}+\frac{%
E_{c}^{\ast }}{2}\dot{\eta}_{1}\dot{\xi}_{2c}.  \notag
\end{eqnarray}%
In Eq. (\ref{quadraVA}) the coordinates of particle $2$ appear evaluated in
either the retarded or the advanced unperturbed light-cone, as indicated by
the subindex $c$. The coefficient of each normal-form binary \ is a function
of $m_{1}$, $m_{2}$ and $\theta $, and is obtained by a Gauge-invariant
combination of derivatives, e.g., the coefficient $B_{21}$ is the same in
both retarded and advanced interactions and given by 
\begin{equation}
B_{21}=i[\frac{\partial ^{2}L_{c}}{\partial \eta _{1}\partial \dot{\xi}_{1}}-%
\frac{\partial ^{2}L_{c}}{\partial \xi _{1}\partial \dot{\eta}_{1}}].
\label{GaugeB}
\end{equation}%
The coefficients were evaluated with a symbolic manipulations software,
starting directly from the effective Lagrangian and taking the necessary
derivatives (in this way we avoid mistakes). The explicit functional
dependences are not given here for brevity, but were checked at the various
limits; Biot-Savart field, Coulomb interaction, Darwin interaction, etc. The
effective Lagrangian for particle 1 is composed of the partial Lagrangians (%
\ref{VAC}), evaluated at the advanced and the retarded light-cones, as in
Eq. (\ref{VAintegr}), plus the kinetic Lagrangian, 
\begin{equation}
L_{eff}^{(1)}=T_{1}-\Gamma L_{+}+(1+\Gamma )L_{-}.  \label{action1}
\end{equation}%
The linearized Euler-Lagrange equation of Lagrangian (\ref{action1}) respect
to $\xi _{1}$ is a linear function of coordinates $\xi _{1}$ , $\eta _{1}$ , 
$\xi _{2+}$ , $\eta _{2+}$ , $\xi _{2-}$ and $\eta _{2-}$, as well as of
their first and second derivatives, i.e., 
\begin{eqnarray}
&&l_{1\xi }(\xi _{1},\eta _{1},\xi _{2+},\eta _{2+},\xi _{2-},\eta _{2-})=
\label{EQqsi1} \\
&&-[(N_{11}+\theta ^{2}G)\xi _{1}+\theta ^{2}G\ddot{\xi}_{1}]+[M_{1}(\ddot{%
\eta}_{1}+2i\dot{\eta}_{1}-\eta _{1})-U_{11}\eta _{1}-iB_{21}\dot{\eta}_{1}]
\notag \\
&&-(R_{+}^{\ast }\xi _{2+}+R_{-}^{\ast }\xi _{2-})-(Y_{+}\dot{\xi}_{2+}+Y_{-}%
\dot{\xi}_{2-})+(T_{+}\ddot{\xi}_{2+}+T_{-}\ddot{\xi}_{2-})  \notag \\
&&-(P_{+}\eta _{2+}+P_{-}\eta _{2-})-(\Lambda _{+}\dot{\eta}_{2+}+\Lambda
_{-}\dot{\eta}_{2-})+(E_{+}\ddot{\eta}_{2+}+E_{-}\ddot{\eta}_{2-}).  \notag
\end{eqnarray}%
The Euler-Lagrange equation of Lagrangian (\ref{action1}) with respect to $%
\eta _{1}$, $\xi _{2}$ and $\eta _{2}$ yields three more linear equations,
which together with Eq. (\ref{EQqsi1}) form a system of four linear delay
equations.

In the following we explain how to include self-interaction into Eq. (\ref%
{EQqsi1}). The Euler-Lagrange equation of the kinetic energy (\ref{quadrat1}%
) with respect to $\xi _{1}$ can be expressed as%
\begin{eqnarray}
&&(1+\gamma _{1}^{2})m_{1}\gamma _{1}(S/C)r_{b}^{3}\Omega ^{2}(\ddot{\eta}%
_{1}+2i\dot{\eta}_{1}-\eta _{1})  \label{kinetdP1} \\
&=&r_{b}^{2}(S/C)\frac{d}{dt}(m_{1}\gamma _{1}r_{b}\dot{u}_{1}),  \notag
\end{eqnarray}%
where we replaced$\ \theta ^{2}$ by $r_{b}^{2}\Omega ^{2}$ into the
definition of $M_{1}$ given below Eq. (\ref{quadrat1}). On the second line
of Eq. (\ref{kinetdP1}) we recognize the variation of the complex momentum, $%
m_{1}\gamma _{1}r_{b}\dot{u}_{1}=p_{1x}+ip_{1y}$, multiplied by the factor $%
r_{b}^{2}(S/C)$. The variation of momentum is the force, so that to account
for self-interaction we add to Eq.(\ref{EQqsi1}) the $x$-component of force (%
\ref{LDE}) multiplied by the factor $r_{b}^{2}(S/C)$, plus $i$ times the $y$%
-component of force (\ref{LDE}) multiplied by $\ r_{b}^{2}(S/C)$, i.e., 
\begin{equation}
r_{b}^{2}(S/C)\frac{2}{3}(1+2\Gamma )r_{b}\frac{d^{3}u_{1}}{dt^{3}}\simeq 
\frac{2\theta ^{3}}{3}(1+2\Gamma )\dddot{u}_{1},  \label{LDEqsi1}
\end{equation}%
where the dots represent derivative respect to scaled time $\tau \equiv
\Omega t$. Using Eq. (\ref{LDEqsi1}) with $u_{1}$ given by Eq. (\ref%
{coordinates}), and expanding up to the linear order, yields the order-zero
offensive force (\ref{offending}) plus the linearization of the self-force
in gyroscopic coordinates. The offensive force (\ref{offending}) is the
nonhomogeneous term of the variational equation, and shall be dealt with in
Appendix C. In this Section we discard it and keep only the linear part of
the variational equations. Notice that Eq. (\ref{kinetdP1}) came out
naturally in the form of variation of momentum multiplied by $r_{b}^{2}$.
This instructive normalization suggests that we scale the equations of
motion with the size of the unperturbed Coulombian attraction, $1/r_{b}^{2}$%
. This normalization is useful when discussing orders of magnitude, and it
is used in Appendix C to discuss estimates with an intuition about physical
orders of magnitude.

The linear variational equations for $\xi _{1}$, $\eta _{1}$, $\xi _{2}$ and 
$\eta _{2}$ form a set of four linear delay equations, a system that can be
solved in general by Laplace transform \cite{Bellman}. In the following we
focus on the planar normal-mode solutions of this linear system, with a 
\emph{fast} ping-pong frequency $\ \lambda _{xy}\Omega /\theta $, a
definition motivated by Eq. (\ref{defstiff}). The complex number $\lambda
_{xy}$ is so far arbitrary, but a \emph{harmonic} oscillation is defined by
an imaginary $\lambda _{xy}$. In the following we substitute $\xi _{1}=A\exp
(\lambda _{xy}\Omega t/\theta )$, $\eta _{1}=B\exp (\lambda _{xy}\Omega
t/\theta )$, $\xi _{2}=C\exp (\lambda _{xy}\Omega t/\theta )$ and $\eta
_{2}=D\exp (\lambda _{xy}\Omega t/\theta )$ into the linearized equations
and assume $\theta $ small and $|\lambda _{xy}|$ of order-one or larger, as
discussed below Eq. (\ref{defstiff}). This yields four homogeneous linear
equations for $A$, $B$, $C$ and $D$, and a nontrivial solution exists only
if the determinant vanishes. Using a symbolic manipulations software this
determinant evaluates to 
\begin{eqnarray}
&&1-\frac{2(1+2\Gamma )\theta ^{2}\lambda _{xy}}{3}+\frac{(1+2\Gamma )\theta
^{4}\lambda _{xy}^{2}}{9}-\frac{2}{27}\frac{\mu }{M}(1+2\Gamma )^{3}\theta
^{6}\lambda _{xy}^{3}+\cdots  \label{detXY} \\
&&+\frac{\mu \theta ^{4}}{M}(1+\frac{7}{\lambda _{xy}^{2}}+\frac{5}{\lambda
_{xy}^{4}})[(1+2\Gamma )\sinh (2\lambda _{xy})-2(1+2\Gamma +2\Gamma
^{2})\cosh ^{2}(\lambda _{xy})]  \notag \\
&&-2\frac{\mu \theta ^{4}}{M}(\frac{1}{\lambda _{xy}}+\frac{5}{\lambda
_{xy}^{3}})[2(1+2\Gamma )\cosh ^{2}(\lambda _{xy})-(1+2\Gamma +2\Gamma
^{2})\sinh (2\lambda _{xy})]=0,  \notag
\end{eqnarray}%
where $M\equiv m_{1}+m_{2}$. In our unit system, the total mass of hydrogen
is $M=1825$. For Dirac's retarded-only electrodynamics, Eq. (\ref{detXY})
with $\Gamma =0$, we obtain the following planar normal-mode condition for
ping-pong modes (the stiff-limit) 
\begin{eqnarray}
&&(1+\frac{2}{\lambda _{xy}}+\frac{7}{\lambda _{xy}^{2}}+\frac{10}{\lambda
_{xy}^{3}}-\frac{5}{\lambda _{xy}^{4}}+\cdots )(\frac{\mu \theta ^{4}}{M}%
)\exp (-2\lambda _{xy})  \label{DiracXY} \\
&=&1-\frac{2}{3}\theta ^{2}\lambda _{xy}+\frac{1}{9}\theta ^{4}\lambda
_{xy}^{2}+\cdots ,  \notag
\end{eqnarray}%
Comparing Eq. (\ref{DiracXY}) to Eq. (\ref{DiracZ}) of Appendix B we find
that the quasi-degeneracy phenomenon exists only in three cases; (i) $\Gamma
=0$, Dirac's theory with retarded-only interactions, (ii) $\Gamma =-1$, a
non-physical advanced-only interactions theory, and (iii) $\Gamma =-1/2$ ,
the dissipative Fokker theory of Ref.\cite{dissipaFokker} and the
action-at-a-distance electrodynamics\cite{astar2B}. This discriminating
degeneracy is an interesting curiosity, and in this paper we disregard the
two other dynamics that exhibit the quasi-degeneracy phenomenon, (ii) and
(iii). \ In the next Section we study a three-frequency orbit of the
hydrogen atom in Dirac's electrodynamics with retarded-only fields ($\Gamma
=0$ ), the physically sound choice to describe hydrogen in nature.

\section{The three-frequency orbit}

\bigskip

As discussed in Section IV, in Dirac's electrodynamics there is a remarkable
quasi-degeneracy of the perpendicular and the planar tangent dynamics. In
the large-$|\lambda |$ limit, both Eq. (\ref{DiracXY}) and Eq. (\ref{DiracZ}%
) of Appendix B reduce to

\begin{equation}
(\frac{\mu \theta ^{4}}{M})\exp (-2\lambda )=1,  \label{Istar}
\end{equation}%
henceforth called the degenerate stiff-limit. The value of $(\mu /M)$ for
hydrogen in Eq. (\ref{Istar}) is a small factor of about $(1/1825)$.
Equation (\ref{Istar}) requires that $\lambda $ has a negative real part
given by $\func{Re}(\lambda )\equiv $ $-\sigma \equiv -\ln (\sqrt{\frac{M}{%
\mu \theta ^{4}}})$. For the atomic magnitude, $\theta ^{-1}\sim 137.036$,
the value of $\sigma $ is about $\sigma \simeq 15.0$. On the other hand, the
imaginary part of $\lambda $ can be an arbitrary multiple of $\pi $, so that
the general solution of Eq. (\ref{Istar}) is 
\begin{equation}
\lambda =-\sigma +\pi qi,  \label{unperastar}
\end{equation}%
where $i\equiv \sqrt{-1}$ and the integrer $q$ is the ping-pong index. \
Notice that the real part of $\lambda $ is always negative, so that the
tangent dynamics about the circular orbit is stable in the stiff-limit. The
unfolding of the degeneracy comes with the terms of order $1/\lambda ^{2}$
and $\theta ^{4}\lambda ^{2}$, as found in Eq. (\ref{DiracXY}) and Eq. (\ref%
{DiracZ}) of Appendix B. The exact roots of Eqs. (\ref{DiracXY}) and (\ref%
{DiracZ}) near the limiting root (\ref{unperastar}) are defined,
respectively, by 
\begin{eqnarray}
\lambda _{xy}(\theta ) &\equiv &-\sigma _{xy}+\pi qi+i\epsilon _{1},
\label{pair} \\
\lambda _{z}(\theta ) &\equiv &-\sigma _{z}+\pi qi+i\epsilon _{2},  \notag
\end{eqnarray}%
where $\epsilon _{1}(\theta )$ and $\epsilon _{2}(\theta )$ are real $%
O(\theta )$ numbers. The three-frequency orbit is formed from an initial
circular orbit as follows; (i) Initially, the self-force (\ref{offending})
dissipates energy, essentially the radiation of the circular orbit. Some of
the radiated energy is absorbed \emph{directly} by the PP oscillations. The
slow guiding-center circular motion may also loose radius to account for
some of the energy loss, spiralling in for a small number of turns. (ii)
After the PP modes absorb enough energy, their amplitudes grow and the PP
oscillations become neutrally stable, which is illustrated in Fig. 1. As
discussed in Appendix C, at a finite PP amplitude, a harmonic solution to
the nonlinear variational equations exists, i.e., the $\sigma ^{\prime }s$
of Eq. (\ref{pair}) vanish, so that the$\ \lambda ^{\prime }s$ become purely
imaginary. This balancing is achieved at relatively small amplitudes, i.e.,
near the circular orbit, and (iii) The radiation of the PP modes starts to
interfere with the orbital radiation if the orbit is a resonant one. In the
following we discuss a surprisingly simple stabilization mechanism that
operates \emph{after} the fast harmonic oscillations are established near
the guiding-center orbit.

We henceforth assume that the three-frequency orbit of Fig. 1 is defined by
the following combination of a planar PP mode and a perpendicular PP mode 
\begin{eqnarray}
x_{k}+iy_{k} &\equiv &r_{b}\exp (i\Omega t)[d_{k}+2\xi _{k}^{\ast }],
\label{XYZPP} \\
x_{k}-iy_{k} &\equiv &r_{b}\exp (-i\Omega t)[d_{k}^{\ast }+2\xi _{k}], 
\notag \\
z_{k} &\equiv &r_{b}Z_{k},  \notag
\end{eqnarray}%
with 
\begin{eqnarray}
\xi _{k} &=&R_{k}^{xy}\exp [(\pi qi+i\epsilon _{1}^{\rho })\Omega t/\theta ],
\label{multiscalesolution} \\
Z_{k} &=&\func{Re}\{R_{k}^{z}\exp [(\pi qi+i\epsilon _{2}^{\rho })\Omega
t/\theta ]\},  \notag
\end{eqnarray}%
where $R_{k}^{xy}$ and $R_{k}^{z}$ are respectively the amplitude of the
planar PP mode and the amplitude of the perpendicular PP mode ( $k=1$ for
the electron and $k=2$\ for the proton). Notice that we chose the same $q$
for the two perpendicular oscillations of Eq. (\ref{multiscalesolution}), so
that $q$ cancels and the difference between the frequencies of \ Eq. (\ref%
{multiscalesolution}) defines a beat in a slow timescale. The far-electric
field of the electron, at a far distance $r_{\infty }$, is obtained by
exchanging indices and disregarding the first term on the right-hand side of
Eq. (\ref{Lienard-E}) of Appendix A,

\begin{equation}
\mathbf{E}_{1}=-\frac{\mathbf{n}\times (\mathbf{n}\times \mathbf{a}_{1-})}{%
(1-\mathbf{n\cdot v}_{1-})^{3}r_{\infty }}+\frac{\mathbf{n\times (v}%
_{1-}\times \mathbf{a}_{1-})}{(1-\mathbf{n\cdot v}_{1-})^{3}r_{\infty }},
\label{far-electric}
\end{equation}%
where we included a minus-one factor to account for the negative electronic
charge. In Eq. (\ref{far-electric}), the unit vector $\mathbf{n}$ points
from the retarded position of the electron to the observation point at
infinity, while $\mathbf{v}_{1-}$ and $\mathbf{a}_{1-}$ are respectively the
Cartesian velocity and acceleration of the electron (particle $1$). The
Poynting flux is proportional to $|\mathbf{E}_{1}|^{2}$, so that the common
denominator $(1-\mathbf{n\cdot v}_{1-})^{3}$on the right-hand side of Eq. (%
\ref{far-electric}) can be factored off $|\mathbf{E}_{1}|$, and is
henceforth ignored. For the radiation of a circular orbit of atomic
magnitude, the second term on the right-hand side of Eq. (\ref{far-electric}%
) is small and can be disregarded. This is the dipole approximation of
atomic physics, valid when the interparticle distance is smaller than the
radiated wavelength. This approximation\emph{\ is not valid} for the
three-frequency orbit of Fig.1 because of the short wavelength of the PP
modes. For the far-field of the three-frequency orbit of Fig. 1, the second
term on the right-hand side of Eq. (\ref{far-electric}) must be kept, and is
henceforth called the spin-radiation field. Because of this quadratic
spin-radiation field, the vector product of the mutually orthogonal PP modes
(\ref{XYZPP}) beats at a slow frequency (cosine times cosine averages to
cosine of the difference). The spin-radiation field of the electron, $%
\mathbf{E}_{PP}$, averaged over the fast timescale and assuming $%
|R_{1}^{z}|\simeq $ $|R_{1}^{xy}|\simeq \rho _{1}$ in Eq. (\ref%
{multiscalesolution}), has the following amplitude 
\begin{equation}
|\mathbf{E}_{PP}|=\frac{\mu \theta ^{2}\rho _{1}^{2}\pi ^{3}q^{3}}{r_{\infty
}}\cos ([1+\frac{(\epsilon _{2}^{\rho }-\epsilon _{1}^{\rho })}{\theta }%
]\Omega t).  \label{pingpongbeat}
\end{equation}%
The oscillation of Eq. (\ref{pingpongbeat}) is determined by the frequency
difference (beat) of the mutually orthogonal PP modes of orbit (\ref%
{multiscalesolution}),%
\begin{equation}
\Delta w_{PP}\equiv \lbrack 1+\frac{(\epsilon _{2}^{\rho }-\epsilon
_{1}^{\rho })}{\theta }]\Omega .  \label{rotation}
\end{equation}%
Because of the quasi-degeneracy property, $\epsilon _{2}^{\rho }$ and $%
\epsilon _{1}^{\rho }$ are $O(\theta )$ quantities, so that $\Delta w_{PP}$
in Eq. (\ref{rotation}) defines\ a beat of the order of the guiding-center
frequency $\Omega $. The first term of the right-hand side of Eq. (\ref%
{far-electric}), averaged over the fast timescale, yields the usual dipole
field $\mathbf{E}_{d}$ of the unperturbed circular motion, with magnitude 
\begin{equation}
|\mathbf{E}_{d}|=\frac{\mu \theta ^{4}}{r_{\infty }}\cos (\Omega t).
\label{dipole}
\end{equation}%
Poynting's theorem gives a necessary condition for $\mathbf{E}_{PP}$ to
destruct $\mathbf{E}_{d}$ by interference, viz., both fields must oscillate
at the same frequency. This\ yields the resonance condition 
\begin{equation}
\Omega =\Delta w_{PP}.  \label{resonancePPC}
\end{equation}%
Notice that at resonance, the three-frequency orbit defined by Eqs. (\ref%
{XYZPP}) and (\ref{multiscalesolution}) has only \emph{two} independent
frequencies\cite{Takens}, because of relation (\ref{resonancePPC}) between
its frequencies. In the gyroscopic coordinates (\ref{multiscalesolution}),
the resonant orbit has only one frequency, and in Appendix C such resonant
orbit is called a harmonic solution. Use of Eq. (\ref{resonancePPC}) with
Eq. (\ref{rotation}) yields 
\begin{equation}
\epsilon _{2}^{\rho }-\epsilon _{1}^{\rho }=0.  \label{resonance1}
\end{equation}%
In the following we approximate $\epsilon _{1}^{\rho }$ and $\epsilon
_{2}^{\rho }$ respectively by the $\epsilon _{1}$ and $\epsilon _{2}$ of the
linear problem, as defined by Eq. (\ref{pair}), assuming that essentially
the real parts of the linear eigenvalues (\ref{pair}) are modified by the
finite-amplitude correction (\ref{stiffdet}), while the imaginary parts
acquire only a small correction. According to this approximation, we expand
the small correction in powers of $\theta $ and take the first term 
\begin{equation}
\epsilon _{2}-\epsilon _{1}\simeq \epsilon _{2}^{\rho }-\epsilon _{1}^{\rho
}+p\theta +\cdots .  \label{approximation}
\end{equation}%
This approximation is justified in Appendix C and the linear coefficient $p$
should depend on the radius of gyration, $\rho _{1}\leq 1/\pi q$. Condition (%
\ref{resonance1}) together with Eqs. (\ref{rotation}) and (\ref{Kepler})
gives $\Omega $ as 
\begin{equation}
\Omega =\mu \theta ^{2}(\epsilon _{2}-\epsilon _{1})+(1-p)\theta ^{3}+\cdots
,  \label{Rydberg-Ritz}
\end{equation}%
a Rydberg-Ritz-like formula. Resonant frequency $\Omega $ is expressed by an
eigenvalue difference in another way; the exact difference between the
frequencies of the orthogonal PP modes, Eq. (\ref{resonancePPC}). The PP
frequencies should be rigorous linear eigenvalues of the tangent dynamics of
the three-frequency orbit, but here we consider only the neighborhood of the
circular orbit, and keep to approximation (\ref{Rydberg-Ritz}).

We henceforth set $p=1$ into Eq. (\ref{approximation}) as a qualitative
approximation, yielding 
\begin{equation}
\epsilon _{1}-\epsilon _{2}+\theta =0,  \label{resonance}
\end{equation}%
which has the solutions listed in Table 1. The root-searching problem of Eq.
(\ref{resonance}) is well posed and for each integer $q$ there is a single
solution $\theta $ \ to Eqs.(\ref{resonance}), (\ref{DiracXY}) and (\ref%
{DiracZ}), i.e., $\theta $ is quantized by the integer $q$. According to
QED, circular Bohr orbits have maximal angular momenta and a radiative
selection rule ( $\Delta l=\pm 1$) restricts the decay from level $k+1$ to
level $k$ only, i.e., circular orbits emit the first line of each
spectroscopic series (Lyman, Balmer, Ritz-Paschen, Brackett, etc.), the
fourth column of Table 1. We have solved Eq. (\ref{resonance}) together with
Eqs. (\ref{DiracXY}) and (\ref{DiracZ}) using a Newton method in the complex 
$\lambda $-plane. The numerically calculated angular momenta $\theta ^{-1}$
are given in Table 1, along with the orbital frequency in atomic units $%
(137^{3}\Omega )/\mu =137^{3}\theta ^{2}(\epsilon _{2}-\epsilon _{1})$, and
the QED first frequency of each spectroscopic series.

\begin{tabular}{|l|l|l|l|l|}
\hline
$k$ & $l_{z}=\theta ^{-1}$ & $137^{3}\theta ^{2}(\epsilon _{2}-\epsilon
_{1}) $ & $w_{QED}$ & $q$ \\ \hline
1 & 185.99 & 3.996$\times $10$^{-1}$ & 3.750$\times $10$^{-1}$ & 7 \\ \hline
2 & 307.63 & 8.831$\times $10$^{-2}$ & 6.944$\times $10$^{-2}$ & 9 \\ \hline
3 & 475.08 & 2.398$\times $10$^{-2}$ & 2.430$\times $10$^{-2}$ & 11 \\ \hline
4 & 577.99 & 1.331$\times $10$^{-2}$ & 1.125$\times $10$^{-2}$ & 12 \\ \hline
5 & 694.77 & 7.667$\times $10$^{-3}$ & 6.111$\times $10$^{-3}$ & 13 \\ \hline
6 & 826.22 & 4.558$\times $10$^{-3}$ & 3.685$\times $10$^{-3}$ & 14 \\ \hline
7 & 973.12 & 2.790$\times $10$^{-3}$ & 2.406$\times $10$^{-3}$ & 15 \\ \hline
8 & 1136.27 & 1.752$\times $10$^{-3}$ & 1.640$\times $10$^{-3}$ & 16 \\ 
\hline
9 & 1316.44 & 1.127$\times $10$^{-3}$ & 1.173$\times $10$^{-3}$ & 17 \\ 
\hline
10 & 1514.40 & 7.403$\times $10$^{-4}$ & 8.678$\times $10$^{-4}$ & 18 \\ 
\hline
11 & 1730.93 & 4.958$\times $10$^{-4}$ & 6.600$\times $10$^{-4}$ & 19 \\ 
\hline
12 & 1966.77 & 3.379$\times $10$^{-4}$ & 5.136$\times $10$^{-4}$ & 20 \\ 
\hline
13 & 2222.70 & 2.341$\times $10$^{-4}$ & 4.076$\times $10$^{-4}$ & 21 \\ 
\hline
\end{tabular}

\ 

Caption to Table 1: Quantum number $k$ of the QED transition $k+1\rightarrow
k$, numerically calculated angular momenta $l_{z}=\theta ^{-1}$ in units of $%
e^{2}/c$, orbital frequencies in atomic units $(137\theta
)^{3}=137^{3}\theta ^{2}(\epsilon _{2}-\epsilon _{1})$, the corresponding $%
k^{th}$circular line of QED in atomic units, $w_{QED}\equiv \frac{1}{2}(%
\frac{1}{k^{2}}-\frac{1}{(k+1)^{2}})$ , and the integer $q$ of Eqs. (\ref%
{pair}) and (\ref{multiscalesolution}).

\bigskip

Table 1 illustrates the fact that a resonance among PP modes predicts
magnitudes precisely in the atomic scale, as first discovered in Ref. \cite%
{dissipaFokker}. In Ref. \cite{dissipaFokker} we had to jump the integer $q$
by twenty units\ for a complete quantitative and qualitative agreement with
the Bohr atom. The qualitative agreement achieved by Table 1 is superior in
this respect; but the correspondence is still not perfect for $q<7$, after
which $q$ increases one by one, in qualitative agreement with QED. The
angular momenta in the second row of Table 1 should be compared with the
QED\ values, i.e., $\theta ^{-1}=137.036k$. In Table 2 we give the anomalous
roots for $q<7$ . Notice that the angular momenta are still of the order of
Planck's constant, but the orbital frequencies do not correspond to any line
of hydrogen. Since the approximation at Eq. (\ref{approximation}) uses an
expansion in $\theta $, we should expect it to fail for the largest values
of $\theta $, as it did.

\begin{tabular}{|l|l|l|}
\hline
$l_{z}=\theta ^{-1}$ & $137^{3}\theta ^{2}(\epsilon _{2}-\epsilon _{1})$ & $%
q $ \\ \hline
93.26 & 3.16 & 1 \\ \hline
51.06 & 1.93 & 2 \\ \hline
57.26 & 1.36 & 3 \\ \hline
76.14 & 5.82 & 4 \\ \hline
103.92 & 2.29 & 5 \\ \hline
140.37 & 0.929 & 6 \\ \hline
\end{tabular}

\bigskip

Caption to Table 2: Numerically calculated angular momenta $l_{z}=\theta
^{-1}$ in units of $e^{2}/c$, orbital frequencies in atomic units $%
(137\theta )^{3}=137^{3}\theta ^{2}(\epsilon _{2}-\epsilon _{1})$, and the
values of the integer $q$ of Eqs. (\ref{pair}) and (\ref{multiscalesolution}%
).

\bigskip

In the following we give a second derivation of Poynting's resonance
condition (\ref{resonancePPC}). The angular momentum of gyration along the
orbital plane, calculated with Eq. (\ref{multiscalesolution}) and averaged
over the fast timescale is%
\begin{eqnarray}
l_{x}+il_{y} &=&\mu r_{b}^{2}|R_{1}^{xy}||R_{1}^{z}|\frac{\pi q\Omega }{%
\theta }b_{1}\exp [i(\epsilon _{2}^{\rho }-\epsilon _{1}^{\rho }+\theta
)\Omega t/\theta ]  \label{angularmomentum} \\
&=&\frac{\pi q}{\theta ^{2}}|R_{1}^{xy}||R_{1}^{z}|\exp [i(\epsilon
_{2}^{\rho }-\epsilon _{1}^{\rho }+\theta )\Omega t/\theta ],  \notag
\end{eqnarray}%
where we used Eqs. (\ref{Kepler}) and (\ref{RB}). The angular momentum
vector of Eq. (\ref{angularmomentum}) rotates at the same frequency $\Delta
w_{PP}$ of Eq. (\ref{rotation}), the ping-pong beat. Since angular momentum
carries inertia, its rotation produces a gyroscopic torque on the orbit, so
that the guiding-center motion should display an oscillation at the
frequency $\Delta w_{PP}$ of Eq. (\ref{rotation}). The derivation of Eq. (%
\ref{angularmomentum}) assumed that the guiding-center orbit is a circular
orbit of frequency $\Omega $, therefore we must have $\Omega =\Delta w_{PP}$%
. In Appendix C we give a third derivation of resonance (\ref{resonancePPC}%
), by averaging the electronic spin-radiation field over the fast timescale.
The resonance is needed for the averaged spin-radiation field to rotate at
the guiding-center frequency, so that it can participate in the
guiding-center dynamics.

\bigskip

\section{Conclusions and Discussion}

We stress that the QED condition, that $\theta $ is of the order of the
fine-structure constant, was never used in the calculations. The PP
frequencies were calculated for \emph{arbitrary} $\theta $, and it was the
resonance condition of Poynting's theorem alone, a \emph{nonradiation
condition}, that calculated $\theta $ in the atomic region. The
quasi-degeneracy phenomenon was already found in Ref. \cite{dissipaFokker},
and we expect it to be a universal property of electromagnetic dynamics with
delay, as follows. The corrections to the limiting form (\ref{Istar}) are
controlled by powers of $1/\lambda $ and $\theta ^{2}\lambda $, so that the
resonance condition essentially poses a relation $\theta ^{2}\lambda \propto
1/\lambda $, \ which yields $\theta \cong c/|\lambda |=c/(\pi q)$, in
agreement with Table 1, with Ref.\cite{dissipaFokker} and with the Bohr atom
of QED\cite{Bohr}. The finite-amplitude corrections discussed in Appendix C
introduce again powers of $\rho \simeq 1/|\lambda |$, so that the
qualitative behavior should persist.

In Refs. \cite{PRL} we argued that a stable orbit should emit at a neutrally
stable frequency of its tangent dynamics. The physical process near the
stable orbit is that the two terms on the right-hand side of Eq. (\ref%
{far-electric}) do not compensate exactly, so that there is a net radiation 
\emph{at} the resonant frequency. Moreover, the guiding motion of the
three-frequency orbit is not exactly a Coulombian circular orbit, because
the perturbative scheme should correct the Coulombian approximation
significantly, specially for the first states, where $\sup (\rho )\simeq
1/\pi q$ \ is larger. The orbital frequencies of Table 1 are only an
approximation to the emitted frequencies, an approximation that should be
worse for the low values of $q$, again in agreement with QED. More
generally, according to Kurtzweil's \emph{small delays don't matter }theorem%
\emph{\ }\cite{Placido}, small \ $\theta $ orbits should be solutions of a
limiting \emph{ordinary} differential equation obtained by setting all
delays to zero. This theorem yields that the limiting orbits should be the
Coulombian orbits. In agreement with this, the gyration radius of Eq. (\ref%
{relativeradius}) is a smaller fraction of the orbital radius for larger
values of $q$ , the $\sup (\rho )\rightarrow 1/\pi q$ limit of Appendix C.
We call this limit the Kurtzweil \emph{correspondence limit}.

For $\theta $ in the atomic magnitude, there is a nontrivial stabilization
mechanism involving PP modes, a mechanism that selects discrete orbits by a
resonance. The ping-pong modes form a channel that can interfere and absorb
the energy radiated by the orbital motion, a dynamics that involves a
resonance between the fast and the slow timescales. A multiscale solution, 
\emph{without assuming} that the guiding-center motion is circular, is still
to be worked out. The multiscale solution involves a general \emph{%
guiding-center }slow motion \emph{plus} a fast PP gyration of small
amplitude. After the fast dynamics is balanced locally, the multiscale
method should yield differential equations for the guiding-center orbit by a
Fredholm alternative\cite{vereador, Mallet-Paret,Politi}. Qualitatively, we
expect the guiding-center equations of motion to accept richer orbits than
illustrated in Figs. 1 and 3. The other possible orbital topologies are: (i)
gyration plane perpendicular to the normal connecting the particles, a
dumbbell-like spinning, (ii) fast gyration restricted to the orbital plane,
and (iii) non-planar guiding-center motion, like a $p$ orbital. The
intuition about the dynamics should follow the general guidelines learnt
from analyzing the circular-guiding-center-three-frequency-orbits of Figs. 1
and 3, as summarized in the following. For a resonant orbit, the electronic
spin-radiation force, averaged over the fast timescale, yields an $O(1)$
slowly rotating force at the proton that perturbs the protonic equilibrium
position along the guiding-center orbit. The protonic attraction cancels the
electronic self-interaction, so that the guiding-center motion does not
dissipate (any guiding-center motion). The averaging of the electronic
spin-radiation force involves a resonance, and this is the reason for
discrete orbits from the perspective of the equations of motion for the
guiding-center. We derived the resonance condition in three different ways
here; (i) In Section V we derived the resonance condition using Poynting's
theorem, (ii) Still in Section V, we derived the same resonance condition
using general arguments about the gyroscopic torque, and last (iii) In
Appendix C we derived the resonance condition from the detailed balancing of
the guiding-center dynamics.

In the process of emission, the sharp line is emitted when the dynamics is
approaching the stable orbit, as discussed in Ref. \cite{PRL}. This long
timescale is to be compared with the time of spontaneous decay of QED for
hydrogen; about $10^{6}$ orbital periods or $10^{-10}$ seconds. On the other
hand, the dynamics has a fast timescale, the frequency of the PP modes, of
about $\pi q/\theta \simeq 1000$ times the orbital frequency. This fast
frequency is about $10^{20}$ Hertz and resonates with the X-ray frequencies
used in the Compton effect \cite{Dodd}, an interesting coincidence. The
ping-pong frequencies of hydrogen have the magnitude of the zitterbewegung
frequency of Dirac's relativistic version of Schroedinger's equation\cite%
{Hestenes}. In modern Aharonov-Bohm experiments \cite{Tonomura}, the
ballistic electron passes at a manometric distance from the electrons inside
the solenoid, such that the PP frequencies, that fall as $1/r$, can be even
at X-rays. The magnetic field interacts with the PP oscillations of the
solenoid electrons, which in turn influences the ballistic electron. This
fact that electrons can play \emph{ping-pong-at-a-distance} with frequencies
up to X-ray suggests the need of an X-ray shielding for the solenoid of the
Aharonov-Bohm experiments, e.g., a layer of lead. \ 

The three-frequency orbit solves several conundrums of the hydrogen atom,
paradoxes that were created by \emph{guessing }that the equations of
classical electrodynamics would accept non-stiff planetary-like orbits at
every scale. The ping-pong phenomenon is a non-trivial feature that is not
present in the ordinary differential equations of planetary dynamics. The
qualitative agreements with QED are listed in the following; (i) the
resonant orbits are quantized by integers that appear naturally because of
the delay, and the orbital frequencies agree reasonably with the Bohr
circular lines. (ii) the angular momenta of the resonant orbits turn out to
be approximate integer multiples of Planck's constant. (iii) The emitted
frequencies are given by a difference of two eigenvalues of a linear
operator, the frequencies of the PP modes, analogously to the Rydberg-Ritz
combinatorial principle of QED. (iv) the averaged angular momentum of
gyration is of the order of the electronic spin-angular-momentum of QED\ 
\cite{Michael,Tomonaga}. The approximation for the gyration angular momentum
is not satisfactory yet though.

We exhibited a new orbit of the hydrogen atom of Dirac's electrodynamics of 
\emph{point }charges with retarded-only fields. The Lyapunov stability of
this orbit should be further investigated. Because we are dealing with a
physical theory, the equations should be sufficiently robust to allow some
kind of numerical integration \cite{Raju}. The stability of the
three-frequency orbit poses a linear set of delay equations, a dynamical
system that needs an initial function as the initial condition, just like
Schroedinger's equation. It would be interesting to learn if this linear
operator has a self-adjoint Fredholm alternative\cite{vereador, Mallet-Paret}%
, like Schroedinger's equation does. The frequency of PP modes is
proportional to the inverse of the interparticle distance, which turns out
to be proportional to the electrostatic potential energy (an accidental
analogy, because these are different phenomena). The potential energy is
thought to be the reason why the $1/r$ appears in Schroedinger's equation;
and it is interesting to notice that a linearized equation for PP modes has
the same generic form. The ping-pong modes appear universally in
electromagnetic many-body dynamics because of the delay, a phenomenon that
has been so far overlooked. The interference mechanism of Poynting's theorem
can produce orbits that do not radiate, if a resonance is satisfied. This
resonance turns out to be satisfied precisely in the atomic magnitude, a
surprise that makes this dynamics interesting for theoretical physics. We
hope that our preliminary findings serve to motivate and guide further
studies of this dynamics, and there is much to be settled yet.

\section{Acknowledgements:}

I thank Savio B. Rodrigues, Guilherme Leal Ferreira, J. C. Egues, Clodoaldo
Ragazzo and Antonio de Toledo Pizza for discussions, Reginaldo Napolitano
for discussions and for reading the manuscript, and the many discussions
with Luigi Galgani, Andrea Carati, Massimo Marino, Antonio Politi, Allan
Lichtenberg and Stefano Ruffo.

\section{\protect\bigskip Appendix A: Electrodynamics of point charges}

In 1947 Eliezer generalized Dirac's covariant subtraction of infinities\cite%
{EliezerReview}. In this generalized electrodynamics of point charges \cite%
{EliezerReview}, the field produced by the point charge is supposed to be
the retarded field plus an intrinsic free field $G$%
\begin{equation}
F_{\mu }^{\nu }=F_{\mu ,ret}^{\nu }+G_{\mu }^{\nu }.  \label{eliG}
\end{equation}%
The free-field $G$ used by Eliezer in Ref. \cite{EliezerReview} is finite
along the particle's trajectory and vanishes when the particle is at rest, 
\begin{equation}
G_{\mu }^{\nu }=\Gamma (F_{\mu ,ret}^{\nu }-F_{\mu ,avd}^{\nu }),  \label{Gk}
\end{equation}%
where $\Gamma $ is a constant \cite{EliezerReview}. This generalized
electromagnetic setting is henceforth called the Eliezer setting (ES). In
Eliezer's theory, the electron and the proton of a hydrogen atom, of charges 
$-1$ and $+1$ respectively, have the following equations of motion \cite%
{EliezerReview}%
\begin{eqnarray}
m_{1}\mathbf{\dot{\upsilon}}_{\mu 1} &=&\frac{2}{3}(1+2\Gamma )[\mathbf{%
\ddot{\upsilon}}_{\mu 1}-||\mathbf{\dot{\upsilon}}_{1}||^{2}\mathbf{\upsilon 
}_{\mu 1}]-[F_{\mu 1,in}^{\nu }+(1+\Gamma )F_{\mu 2,ret}^{\nu }-\Gamma
F_{\mu 2,adv}^{\nu }]\mathbf{\upsilon }_{\nu 1},  \label{Eliezermotion} \\
m_{2}\mathbf{\dot{\upsilon}}_{\mu 2} &=&\frac{2}{3}(1+2\Gamma )[\mathbf{%
\ddot{\upsilon}}_{\mu 2}-||\mathbf{\dot{\upsilon}}_{2}||^{2}\mathbf{\upsilon 
}_{\mu 2}]+[F_{\mu 2,in}^{\nu }+(1+\Gamma )F_{\mu 1,ret}^{\nu }-\Gamma
F_{\mu 1,adv}^{\nu }]\mathbf{\upsilon }_{\nu 2},  \notag
\end{eqnarray}%
where the $\mathbf{\upsilon }_{\mu i}$ stand for the four components ($\mu
=1,2,3,4$) of the Minkowski velocity of particle $i$, double bars stand for
the Minkowski norm and the dot indicates derivative respect to the proper
time of each particle. Equation (\ref{Eliezermotion}) includes an external
field $F_{\mu ,in}^{\nu }$ produced by the other charges of the universe 
\emph{at} each particle's trajectory, which vanishes for the isolated
electromagnetic two-body problem. The first term on the right-hand-side of
Eq. (\ref{Eliezermotion}) is the sourceless combination of half of the
retarded Lienard-Wiechert self-potential minus half of the advanced
Lienard-Wiechert self-potential, all multiplied by $(1+2\Gamma )$ and
evaluated at the particle itself\cite{EliezerReview}. This term generalizes
Dirac's self-interaction force\cite{Dirac}. The last two terms on the
right-hand side of the first line of Eq. (\ref{Eliezermotion}) define the
interaction with the retarded potentials of the proton at the electronic
position, $F_{\mu 2,ret}^{\nu }$, and the interaction with the advanced
potentials of the proton at the electronic position, $F_{\mu 2,adv}^{\nu }$.
The ES has four interesting limits; (i) $\Gamma =0$ is Dirac's
electrodynamics with retarded-only fields \cite{Dirac}, (ii) $\Gamma =-1$ is
a non-physical theory with advanced-only interactions, (iii) $\Gamma =-1/2$
is the action-at-a-distance electrodynamics \cite{FeyWhe} ( notice that the
self-interaction vanishes), and (iv) The limit when $\Gamma $ tends to $-1/2$
from above is the dissipative Fokker setting of Ref. \cite{dissipaFokker},
with a charge renormalization controlled by $(1+2\Gamma )$.

For the non-specialist reader, in the following we write the equations of
motion of Dirac's electrodynamics, $\Gamma =0$, in the intuitive form of
physics textbooks. For the isolated hydrogen atom, the spatial component of
the electronic equation of motion, the first line of Eq. (\ref{Eliezermotion}%
), multiplied by $\sqrt{1-|\mathbf{v}_{1}|^{2}}$, yields%
\begin{equation}
m_{1}\frac{d}{dt_{1}}(\frac{\mathbf{v}_{1}}{\sqrt{1-|\mathbf{v}_{1}|^{2}}})=%
\mathbf{F}_{1}-\mathbf{E}_{2}-\mathbf{v}_{1}\times \mathbf{B}_{2},
\label{Eqfamiliar}
\end{equation}%
where $\mathbf{v}_{1}$ is the Cartesian electronic velocity\cite%
{Anderson,Rohrlich}. The equation for the proton is obtained exchanging the
indices in Eq. (\ref{Eqfamiliar}) and multiplying the last two terms on the
right-hand side of Eq. (\ref{Eqfamiliar}) by minus one, to account for the
positive protonic charge. The first term on the right-hand side of Eq. (\ref%
{Eqfamiliar}) is the force $\mathbf{F}_{1}$ of the electronic fields on the
point electron itself, and is called the Lorentz-Dirac self-force,%
\begin{equation}
\mathbf{F}_{1}=\frac{2}{3}\gamma _{1}^{2}\{\mathbf{\dot{a}}_{1}+\gamma
_{1}^{2}(\mathbf{v}_{1}\cdot \mathbf{\dot{a}}_{1})\mathbf{v}_{1}+3\gamma
_{1}^{2}(\mathbf{v}_{1}\cdot \mathbf{a}_{1})[\mathbf{a}_{1}+\gamma _{1}^{2}(%
\mathbf{v}_{1}\cdot \mathbf{a}_{1})\mathbf{v}_{1}]\},  \label{nonlidissi1}
\end{equation}%
where $\gamma _{1}\equiv 1/\sqrt{1-|\mathbf{v}_{1}|^{2}}$ and $\mathbf{a}_{1}
$ and $\mathbf{\dot{a}}_{1}$ stand for the Cartesian electronic acceleration
and time-derivative of the Cartesian electronic acceleration, respectively
(cf. page 116 of Ref \cite{Rohrlich}). In Eq. (\ref{nonlidissi1}) a dot
between two Cartesian vectors indicates scalar product and a dot over a
vector indicates derivative respect to laboratory time $t_{1}$. The second
term on the right-hand side of Eq. (\ref{Eqfamiliar}) is the electric force,
i.e., the electronic charge, $e=$ $-1$, times the electric field of the
proton acting on the electron, $\mathbf{E}_{2}$. In our unit system with $c=1
$, the retarded electric field of the proton, of charge $+1$, is given by
the Lienard-Wiechert formula\cite{Jackson}%
\begin{equation}
\mathbf{E}_{2}=\frac{\mathbf{n}_{12b}-\mathbf{v}_{2b}}{\gamma _{2b}^{2}(1-%
\mathbf{n}_{12b}\mathbf{\cdot v}_{2b})^{3}r_{12b}^{2}}+\frac{\mathbf{n}%
_{12b}\times \lbrack (\mathbf{n}_{12b}-\mathbf{v}_{2b})\times \mathbf{a}%
_{2b}]}{(1-\mathbf{n}_{12b}\mathbf{\cdot v}_{2b})^{3}r_{12b}},
\label{Lienard-E}
\end{equation}%
$\ $where $\mathbf{v}_{2b}$ and $\mathbf{a}_{2b}$ are respectively the
Cartesian velocity and Cartesian acceleration of the proton at the retarded
time $t_{2b}$ and $\gamma _{2b}\equiv 1/\sqrt{1-|\mathbf{v}_{2b}|^{2}}$. In
Eq. (\ref{Lienard-E}), unit vector $\mathbf{n}_{12b}$ connects the retarded
protonic position to the present electronic position, and $r_{12b}$ is the
interparticle distance along the retarded light-cone. The advanced fields
are obtained by replacing $\mathbf{v}_{2b}$ by $-\mathbf{v}_{2b}$ in Eq. (%
\ref{Lienard-E}), and are not used here, since we deal we Dirac's
retarded-only theory. The third term on the right-hand side of Eq. (\ref%
{Eqfamiliar}) is the magnetic force, i.e., the electronic charge times the
vector product of the electronic velocity $\mathbf{v}_{1}$ by the magnetic
field $\mathbf{B}_{2}$ of the proton. The retarded magnetic field of the
proton is given by the Lienard-Wiechert formula \cite{Jackson}%
\begin{equation}
\mathbf{B}_{2}=\mathbf{n}_{12b}\times \mathbf{E}_{2}.  \label{magnetic-B}
\end{equation}%
For the advanced magnetic field, formula (\ref{magnetic-B}) includes a minus
sign, so that the Poynting flux $\mathbf{E}_{2}\times \mathbf{B}_{2}$
calculated with the purely advanced fields of an\emph{\ unperturbed circular
orbit }is an incoming energy flux. The combination $-(\mathbf{E}_{2}+\mathbf{%
v}_{1}\times \mathbf{B}_{2})$ is usually called the Lorentz force, not to be
confused with the Lorentz-Dirac self-force (\ref{nonlidissi1}). The electric
and magnetic fields (\ref{Lienard-E}) and (\ref{magnetic-B}) depend on the
retarded position, velocity and acceleration of the proton, and this is
where delay enters in the dynamics. Equation (\ref{nonlidissi1}) can be
rearranged as 
\begin{equation}
\mathbf{F}_{1}=\frac{2}{3}\gamma _{1}^{2}U_{1}\{\mathbf{\dot{a}}_{1}+3\gamma
_{1}^{2}(\mathbf{v}_{1}\cdot \mathbf{a}_{1})\mathbf{a}_{1}\},  \label{Savio}
\end{equation}%
where we introduced the dyadic matrix $U_{1}\equiv I+\gamma _{1}^{2}\mathbf{v%
}_{1}\mathbf{v}_{1}^{t}$, a non-singular matrix whose inverse is $%
U_{1}^{-1}=I-\mathbf{v}_{1}\mathbf{v}_{1}^{t}$. The left-hand side of Eq. (%
\ref{Eqfamiliar}), i.e., the variation of the momentum, can be expressed
using $U_{1}$ as%
\begin{equation}
m_{1}\frac{d}{dt_{1}}(\frac{\mathbf{v}_{1}}{\sqrt{1-|\mathbf{v}_{1}|^{2}}}%
)=m_{1}\gamma _{1}U_{1}\mathbf{a}_{1}.  \label{U-force}
\end{equation}%
Using Eqs. (\ref{Savio}) and (\ref{U-force}), equation of motion (\ref%
{Eqfamiliar}) can be expressed as%
\begin{equation}
m_{1}\mathbf{a}_{1}=\frac{2}{3}\gamma _{1}\{\mathbf{\dot{a}}_{1}+3\gamma
_{1}^{2}(\mathbf{v}_{1}\cdot \mathbf{a}_{1})\mathbf{a}_{1}\}-\sqrt{1-|%
\mathbf{v}_{1}|^{2}}U_{1}^{-1}(\mathbf{E}_{2}+\mathbf{v}_{1}\times \mathbf{B}%
_{2}),  \label{newton}
\end{equation}%
where $\mathbf{E}_{2}$ and $\mathbf{B}_{2}$ are the electric and magnetic
fields of the proton, respectively, and we have cancelled the invertible
matrix $U_{1}$ and a power of $\gamma _{1}$. Equation (\ref{newton}) has a
familiar Newtonian-like form, but it is still the full relativistic
equation. In the low-velocity limit, the first term of the right-hand side
of Eq. (\ref{newton}) reduces to the third derivative of the position
multiplied by $2/3$, which is called the Abraham-Lorentz-Dirac self-force.

There exists an action formalism for the Lorentz-force sector of Eq. (\ref%
{Eqfamiliar}), i.e., Eq. (\ref{Eqfamiliar}) without the first term on the
right-hand side. We give a general action including the parameter $\Gamma $
solely for the calculations and crosschecking of Sections III and IV. A
reader interested only in Dirac's theory can set $\Gamma =0$. This action
involves only the instantaneous position and velocity of particle $1$, and
is composed of a kinetic term plus the following electromagnetic action, 
\begin{equation}
\Theta =-\Gamma \int \frac{(1-\mathbf{v}_{1}\cdot \mathbf{v}_{2a})}{%
r_{12a}(1+\mathbf{n}_{12a}\cdot \mathbf{v}_{2a})}dt_{1}+(1+\Gamma )\int 
\frac{(1-\mathbf{v}_{1}\cdot \mathbf{v}_{2b})}{r_{12b}(1-\mathbf{n}%
_{12b}\cdot \mathbf{v}_{2b})}dt_{1}.  \label{VAintegr}
\end{equation}%
In Eq. (\ref{VAintegr}), $\mathbf{v}_{1}$ stands for the Cartesian velocity
of particle $1$ at time $t_{1}$, while $\mathbf{v}_{2a}$ and $\mathbf{v}%
_{2b} $ stand for the Cartesian velocities of particle $2$ respectively at
the advanced time $t_{2a\text{ }}$and at the retarded time $t_{2b}$. The
vector $\mathbf{n}_{12a}$ in Eq. (\ref{VAintegr}) is a unit vector
connecting the advanced position of particle $2$ to the position of particle 
$1$ at its present time $t_{1}$, while unit vector $\mathbf{n}_{12b}$
connects the retarded position of particle $2$ to the present position of
particle $1$ at time $t_{1}$. Still in Eq. (\ref{VAintegr}), $r_{12a}$ and $%
r_{12b}$ indicate the interparticle distance along the advanced and retarded
light-cones, respectively. To derive the equations of motion of particle $1$%
, one needs to add the usual kinetic action to Eq. (\ref{VAintegr}), $%
K_{1}\equiv -\dint m_{1}\sqrt{1-|\mathbf{v}_{1}|^{2}}dt_{1}$ ( the integral
of the kinetic Lagrangian). In Ref.\cite{Anderson} it is shown that formal
minimization of the sum of action (\ref{VAintegr}) with the kinetic action
yields the equations of motion of particle $1$ suffering the electromagnetic
fields of particle $2$, i.e., Eq. (\ref{Eqfamiliar}) without the
self-interaction term. Each integrand of the right-hand side of Eq. (\ref%
{VAintegr}) is a familiar electromagnetic Lagrangian 
\begin{equation}
L_{c}\equiv \frac{(1-\mathbf{v}_{1}\cdot \mathbf{v}_{2c})}{r_{12c}(1+\frac{%
\mathbf{n}_{12}\cdot \mathbf{v}_{2c}}{c})}\equiv -(V_{c}-\mathbf{v}_{1}\cdot 
\mathbf{A}_{c}),  \label{VAC}
\end{equation}%
where $V_{c}$ and $\mathbf{A}_{c}$ are the Lienard-Wiechert scalar potential
and the Lienard-Wiechert vector potential, respectively. We introduced the
quantity $c=\pm 1$ in Eq. (\ref{VAC}) to indicate if the interaction is
along the advanced or the retarded light-cone. Equation (\ref{VAC}) with $%
c=1 $ indicates that particle $2$ is in the future, while $c=-1$ indicates
that particle $2$ is in the past. The quantities of particle $2$ in Eq. (\ref%
{VAC}) \ are to be evaluated at the time $t_{2c}$ defined implicitly by 
\begin{equation}
t_{2c}=t_{1}+\frac{r_{12c}}{c},  \label{lightcone}
\end{equation}%
where $r_{12+}$ is the distance along the advanced light-cone and $r_{12-}$
is the distance along the retarded light-cone. According to Eq. (\ref%
{lightcone}), the time lag along the advanced light-cone is $r_{12+}$ and
the time lag along the retarded light-cone is $-r_{12-}$.

The shortest way to obtain the equations of motion of Dirac's
electrodynamics in any given coordinate system is to start from the
Euler-Lagrange equations of action (\ref{VAintegr}) plus the relativistic
kinetic action. This yields the dynamics without self-interaction of each
particle suffering the electromagnetic fields of the other particle. The
self-force can be added to the Euler-Lagrange equations, watching carefully
for the correct multiplicative factor. The stiff limit of PP modes is
determined by the largest-order derivative appearing in the linearized
equations of motion. In this limit, the contribution of the self-interaction
to the linearized dynamics about a circular orbit is given by the
Abraham-Lorentz-Dirac self-force%
\begin{equation}
\mathbf{F}_{rad}=\frac{2}{3}(1+2\Gamma )\mathbf{\dot{a}}\text{.}  \label{LDE}
\end{equation}%
The contribution of the other nonlinear terms, that become important at a
finite distance from the circular orbit, is discussed in Appendix C. The
electrodynamics of point charges is discussed at length in Refs. \cite%
{EliezerReview,Rohrlich}, while the Lienard-Wiechert potentials,
Lienard-Wiechert fields, Poynting 's theorem and the physics of
electrodynamics is found in innumerous textbooks, e.g. Refs.\cite%
{Anderson,Jackson}.

\bigskip

\section{\protect\bigskip Appendix B: Variational equations along the $\hat{z%
}$ direction}

In this appendix we derive the linearized variational equations for
displacements perpendicular to the orbital plane, henceforth called the $%
\hat{z}$-direction. Since $z_{k}=0$ is an exact solution of the equations,
in this Section we are doing Lyapunov stability. The variational dynamics
along the $z$-direction is uncoupled from the planar variational equations
up to the linear order. In the same way of Section IV, we expand to second
order the implicit light-cone condition and action (\ref{VAintegr}). The
Cartesian coordinates of a transversely perturbed circular orbit are defined
as%
\begin{eqnarray}
x_{k}+iy_{k} &\equiv &r_{b}d_{k}\exp (i\Omega t),  \label{Zperturb} \\
x_{k}-iy_{k} &\equiv &r_{b}d_{k}^{\ast }\exp (-i\Omega t),  \notag \\
z_{k} &\equiv &r_{b}Z_{k},  \notag
\end{eqnarray}%
where $k=1$ for the electron and $k=2$ \ for the proton, $Z_{k}$ is the
transverse perturbation, $d_{1}\equiv b_{1}$ and $d_{2}\equiv -b_{2}$ are
defined in Eq. (\ref{defradius}) and $\Omega $ is the orbital frequency (\ref%
{Kepler}). In the following we calculate the delay function $\varphi _{c}$
of (\ref{light-cone}) by expanding the light-cone time $t_{2c}$ about the
constant lag $r_{b}$ up to the second order in $\ Z_{1}$ and $Z_{2}$. The
distance $r_{12c}$ in Eq. (\ref{lightcone}) is evaluated from the position
of particle $1$ at time $t_{1}$ to the position (\ref{Zperturb}) of particle 
$2$ \ at time $t_{2c}$. Using $t_{2c}$ defined by Eq. (\ref{lightcone}) and
orbit (\ref{Zperturb}), this implicit distance $r_{12c}=|t_{2c}-t_{1}|$ \
evaluates to%
\begin{equation}
r_{12c}^{2}\equiv r_{b}^{2}(1+\phi
)^{2}=r_{b}^{2}[b_{1}^{2}+b_{2}^{2}+2b_{1}b_{2}\cos (\varphi +\theta
c)]+r_{b}^{2}(Z_{1}-Z_{2c})^{2},  \label{lightconeZ}
\end{equation}%
where we expressed $\varphi _{c}$ in terms of the scaled function $\phi $ of
(\ref{scaledphi}). The $Z$ variations decuple from the planar variations
because powers of $Z$ always appear squared, so that there is no mixed
linear term of $Z$ times a linear perturbation of the planar coordinate in
Eq. (\ref{lightconeZ}). Expanding Eq. (\ref{lightconeZ}) up to the second
order and rearranging yields

\begin{equation}
\phi ^{2}+2S\phi =(Z_{1}-Z_{2c})^{2}.  \label{quadraZ}
\end{equation}%
Equation (\ref{quadraZ}) is a quadratic equation for $\phi $, and the
regular solution is given, up to second order, by%
\begin{equation}
\phi =\frac{1}{2S}(Z_{1}-Z_{2c})^{2}.  \label{PHIZ}
\end{equation}%
The coordinate $Z_{2}$ appears evaluated at the advanced/retarded time in
Eq. (\ref{PHIZ}), and to obtain the action up to quadratic terms it is
sufficient to keep the first term $Z_{2}(\tau _{1}+c\theta +\varphi )$ $%
\simeq Z_{2}(\tau _{1}+c\theta )\equiv Z_{2c}$. Using orbit (\ref{Zperturb})
to calculate the numerator of Lagrangian (\ref{VAC}) yields%
\begin{eqnarray}
(1-\mathbf{v}_{1}\cdot \mathbf{v}_{2c}) &=&1+\theta ^{2}\cos (\theta
+c\varphi )b_{1}b_{2}-\theta ^{2}\dot{Z}_{1}\dot{Z}_{2c}\approx  \label{h2Z}
\\
&&C-\theta ^{2}(S-1)\phi -\theta ^{2}\dot{Z}_{1}\dot{Z}_{2c},  \notag
\end{eqnarray}%
while the denominator of Lagrangian (\ref{VAC}) evaluates to%
\begin{equation}
r_{12c}(1+\frac{\mathbf{n}_{12c}\cdot \mathbf{v}_{2c}}{c})=r_{b}[1+\phi
+\theta cb_{1}b_{2}\sin (\theta c+\varphi _{c})+\theta c(Z_{1}-Z_{2c})\dot{Z}%
_{2c}].  \label{h4Z}
\end{equation}%
Notice that the quadratic term $Z_{2c}\dot{Z}_{2c}$ on the right-hand side
of Eq. (\ref{h4Z}) can be dropped because it represents an exact Gauge that
does not affect the Euler-Lagrange equations. We also expand the argument of
the sine function of the right-hand side of Eq. (\ref{h4Z}) until the linear
term in $\varphi _{c}$, so that the quadratic approximation to Eq. (\ref{h4Z}%
) is 
\begin{equation}
r_{12c}(1+\frac{\mathbf{n}_{12c}\cdot \mathbf{v}_{2c}}{c})\approx
r_{b}[S+C\phi +\theta cZ_{1}\dot{Z}_{2c}],  \label{h4ZGauge}
\end{equation}%
where the equivalence sign $\approx $ henceforth means equivalent up to a
Gauge term of second order. Even if a quadratic Gauge term appears in the
denominator, in an expansion up to quadratic order it would still produce a
Gauge and therefore it can be dropped directly from the denominator. The
Lagrangian of action (\ref{VAintegr}) expanded up to second order is 
\begin{eqnarray}
\pounds _{\Gamma } &\approx &(\frac{C}{r_{b}S})\{(1+\Gamma )[1-\theta
^{2}CS^{2}\dot{Z}_{1}\dot{Z}_{2-}-\frac{C^{2}S}{2}(Z_{1}-Z_{2-})^{2}+\theta
C^{2}SZ_{1}\dot{Z}_{2-}]  \label{VAZ} \\
&&-\Gamma \lbrack 1-\theta ^{2}CS^{2}\dot{Z}_{1}\dot{Z}_{2+}-\frac{C^{2}S}{2}%
(Z_{1}-Z_{2+})^{2}-\theta C^{2}SZ_{1}\dot{Z}_{2+}]\}.  \notag
\end{eqnarray}%
We henceforth disregard $O(\theta ^{2})$ corrections, and substitute $C=S=1$
in the coefficients of \ Lagrangian (\ref{VAZ}). Last, we need the kinetic
Lagrangian along orbit (\ref{Zperturb}), i.e., 
\begin{equation}
T_{1}=-m_{1}\sqrt{1-v_{1}^{2}}=-\frac{m_{1}}{\gamma _{1}}\sqrt{1-\gamma
_{1}^{2}\theta ^{2}\dot{Z}_{1}^{2}},  \label{kineticZ}
\end{equation}%
where the dot means derivative with respect to the scaled time $\tau $, $%
\gamma _{1}^{-1}\equiv \sqrt{1-v_{1}^{2}}$ , and we have used that $\Omega
r_{b}=\theta $. Equation (\ref{kineticZ}) expanded up to the second order is%
\begin{equation}
T_{1}=(\frac{1}{r_{b}})\{\frac{-r_{b}m_{1}}{\gamma _{1}}+\frac{\epsilon _{1}%
}{2}\dot{Z}_{1}^{2}+\cdots \},  \label{expaE}
\end{equation}%
where $\epsilon _{1}$ $\equiv r_{b}^{3}m_{1}\gamma _{1}\Omega ^{2}$. Using
Eqs. (\ref{Kepler}) and (\ref{RB}), we obtain $\epsilon _{1}=m_{1}/\mu
=M/m_{2}$. The equation of motion for particle $1$, without the
self-interaction term, is the Euler-Lagrange equation of the quadratic
Lagrangian%
\begin{equation}
L_{eff}^{(1)}=T_{1}+\pounds _{\Gamma }.  \label{LagrangeZ1}
\end{equation}%
The Lagrangian sector of the equation for $Z_{1}$ is%
\begin{equation}
\epsilon _{1}\ddot{Z}_{1}=-[Z_{1}-(1+\Gamma )Z_{2-}+\Gamma Z_{2+}]+\theta
\lbrack \Gamma \dot{Z}_{2+}+(1+\Gamma )\dot{Z}_{2-})+\theta ^{2}[\Gamma 
\ddot{Z}_{2+}-(1+\Gamma )\ddot{Z}_{2-}].  \label{EQZ1}
\end{equation}%
Notice that the left-hand side of Eq. (\ref{EQZ1}) can be written as 
\begin{equation}
\epsilon _{1}\ddot{Z}_{1}=r_{b}^{3}m_{1}\gamma _{1}\Omega ^{2}\ddot{Z}%
_{1}=r_{b}^{2}\frac{dp_{z}}{dt},  \label{add1}
\end{equation}%
which is the force along the $z$-direction multiplied by $r_{b}^{2}$. As
explained above Eq. (\ref{LDEqsi1}), to account for self-interaction we must
add to Eq. (\ref{EQZ1}) the Abraham-Lorentz-Dirac self-force (\ref{LDE})
multiplied by $r_{b}^{2}$ , 
\begin{equation}
r_{b}^{2}\mathbf{F}_{rad}=\frac{2}{3}(1+2\Gamma )\theta ^{3}\dddot{Z}_{1}.
\label{dissi1}
\end{equation}%
The full linearized variational equation for $Z_{1}$ is%
\begin{eqnarray}
\epsilon _{1}\ddot{Z}_{1} &=&\frac{2}{3}(1+2\Gamma )\theta ^{3}\dddot{Z}%
_{1}-[Z_{1}+\Gamma Z_{2+}-(1+\Gamma )Z_{2-}]  \label{DFEQZ1} \\
&&+\theta \lbrack \Gamma \dot{Z}_{2+}+(1+\Gamma )\dot{Z}_{2-}]+\theta
^{2}[\Gamma \ddot{Z}_{2+}-(1+\Gamma )\ddot{Z}_{2-}].  \notag
\end{eqnarray}%
The linearized equation for $Z_{2}$ is obtained by interchanging $Z_{1\text{ 
}}$by $Z_{2}$ and $\epsilon _{1}$ by $\epsilon _{2}$ in Eq. (\ref{DFEQZ1}).
Comparing Eq. (\ref{DFEQZ1}) to Eq. (30) of Ref. \cite{dissipaFokker} we
find that Eqs. (29) and (30) of Ref.\cite{dissipaFokker} are both missing a $%
\theta ^{3}$ factor in front of the $\dddot{Z}_{1}$ term, which is a typo.
After Eq. (30), the other equations of Ref.\ \cite{dissipaFokker} have the
self-interaction included correctly.

A ping-pong normal mode is obtained by substituting $Z_{1}=A\exp (\lambda
_{z}\Omega t/\theta )$ and $Z_{2}=B\exp (\lambda _{z}\Omega t/\theta )$ into
Eq. (\ref{DFEQZ1}) and the corresponding linearized equation for the proton.
Again we use a general $\lambda _{z}$, but a harmonic solution needs a
purely imaginary $\lambda _{z}.$ Setting the determinant to zero yields%
\begin{equation}
\left\vert 
\begin{array}{cc}
1+\frac{M\lambda _{z}^{2}}{m_{2}\theta ^{2}}-\frac{2}{3}(1+2\Gamma )\lambda
_{z}^{3} & G(\theta ,\lambda _{z}) \\ 
G(\theta ,\lambda _{z}) & 1+\frac{M\lambda _{z}^{2}}{m_{1}\theta ^{2}}-\frac{%
2}{3}(1+2\Gamma )\lambda _{z}^{3}%
\end{array}%
\right\vert =0,  \label{detZ}
\end{equation}%
where $G(\theta ,\lambda _{z})\equiv \lbrack 1-(1+2\Gamma )\lambda
_{z}-\lambda _{z}^{2}]\cosh (\lambda _{z})-[(1+2\Gamma )-\lambda
_{z}-(1+2\Gamma )\lambda _{z}^{2}]\sinh (\lambda _{z})$. The stiff-limit is
when $|\lambda _{z}|$ is large, and we should keep in mind that the
hyperbolic functions in $G(\theta ,\lambda _{z})$ can acquire a large
magnitude \cite{astar2B}. Multiplying determinant (\ref{detZ}) by $\mu
\theta ^{4}/(M\lambda _{z}^{4})$ we obtain%
\begin{eqnarray}
&&1-\frac{2}{3}(1+2\Gamma )\theta ^{2}\lambda _{z}+\frac{4}{9}\frac{\mu }{M}%
\theta ^{4}\lambda _{z}^{2}  \label{EQ5Z} \\
&&-\frac{\mu \theta ^{4}}{M}\{[1-\frac{1}{\lambda _{z}^{2}}+\frac{(1+2\Gamma
)}{\lambda _{z}}]\cosh (\lambda _{z})-[\frac{1}{\lambda _{z}}+(1+2\Gamma )(1-%
\frac{1}{\lambda _{z}^{2}})]\sinh (\lambda _{z})\}^{2}=0,  \notag
\end{eqnarray}%
up to small $O(\theta ^{2})$ terms. The stiff-mode condition defined by Eq. (%
\ref{EQ5Z}) with $\Gamma =-1/2$ is equation (33) of Ref.\cite{dissipaFokker}%
, i.e.,%
\begin{equation}
1-\frac{2}{3}\theta ^{2}\lambda _{z}+\frac{4\mu }{9M}\theta ^{4}\lambda
_{z}^{2}-\frac{\mu \theta ^{4}}{M}[(1-\frac{1}{\lambda _{z}^{2}})[\cosh
^{2}(\lambda _{z})-\frac{1}{\lambda _{z}}\sinh (2\lambda _{z})]^{2}=0.
\label{fourthZ}
\end{equation}%
In Ref \cite{dissipaFokker} there is another typo in passing from Eq. (33)
to Eq. (34); Eq. (34) is missing a bracket that should start after the $(\mu
\theta ^{4}/M)$ factor and close at the end of Eq. (34). The stiff-limit in
Dirac's theory with retarded-only fields, $(\Gamma =0)$, is 
\begin{equation}
1-\frac{2}{3}\theta ^{2}\lambda _{z}+\frac{4\mu }{9M}\theta ^{4}\lambda
_{z}^{2}-\frac{\mu \theta ^{4}}{M}[1+\exp (-2\lambda _{z})](1+\frac{2}{%
\lambda _{z}}-\frac{1}{\lambda _{z}^{2}}-\frac{1}{\lambda _{z}^{3}}+\frac{1}{%
\lambda _{z}^{4}})=0,  \label{DiracZ}
\end{equation}%
and the appearance of the negative exponential is related to the
retardation-only. Comparing Eq. (\ref{EQ5Z}) to Eq. (\ref{detXY}) we find
that the quasi-degeneracy phenomenon exists only for (i) $\Gamma =0$ , i.e.,
Dirac's theory, (ii) $\Gamma =-1$, a non-physical advance-only case, and
(iii) $\Gamma =-1/2$ , the action-at-a-distance electrodynamics\cite{FeyWhe}
and the dissipative Fokker dynamics of Ref. \cite{dissipaFokker}.

\bigskip

\section{\protect\bigskip Appendix C: The ping-pong solutions}

In this Appendix we discuss the existence of a harmonic orbit such as given
by Eq. (\ref{multiscalesolution}) and illustrated in Figs. 1 and 3. In the
following we show that an orbit such as (\ref{multiscalesolution}) exists in
the limit where $\rho _{k}\equiv \sup (|R_{k}^{z}|,|R_{k}^{xy}|)\rightarrow 
\frac{1}{|\lambda |}$, a limit where the stiff gyration approaches the speed
of light. In this limit, the dominant field of the particle is the
far-electric field, Eq. (\ref{far-electric}). The near-electric field is the
first term on the right-hand side of Eq. (\ref{Lienard-E}), which vanishes
because $\gamma _{k}\rightarrow \infty $. The far-magnetic interaction is
the next-to-leading term in size, and is disregarded in the following. We
also disregard the self-interaction, Eq. (\ref{nonlidissi1}), because its
contribution to the force \emph{normal}\textbf{\ }to the velocity is
smaller. The equation of motion for the proton is obtained by exchanging
indices in Eq. (\ref{newton}), and the leading fast dynamics \emph{normal}
to the fast velocity is%
\begin{equation}
m_{2}\mathbf{a}_{2}=-\sqrt{1-|\mathbf{v}_{2}|^{2}}U_{2}^{-1}\{\frac{\mathbf{%
n\times \lbrack (n\times \dot{v}}_{1-})-(\mathbf{v}_{1-}\times \mathbf{\dot{v%
}}_{1-})]}{(1-\mathbf{n\cdot v}_{1-})^{3}r_{21b}}\},  \label{stiffE}
\end{equation}%
where underscore minus indicates that particle $1$ is in the past
light-cone. The unit vector $\mathbf{n}$ points from the retarded position
of particle $1$, which is not indicated with underscore to avoid an
overloaded notation. For the PP modes, the fact that the electronic
coordinates on the right-hand side of Eq. (\ref{stiffE}) are evaluated at $%
t_{1-}\simeq t_{2}-r_{b}$ makes \emph{a lot }of difference, because the PP
modes execute complete periods during this time-lag. Since the gyration
amplitude is small compared with $r_{b}$, we henceforth assume that the
distance $r_{21b}$ along the retarded light-cone on the right-hand side of
Eq. (\ref{stiffE}) is constant and given by $r_{b}$. Along the orbit of Fig.
1, the fast gyration can be parallel to the normal, so that $(1-\mathbf{n}%
\cdot \mathbf{v}_{1-})^{3}$ becomes arbitrarily small, and we henceforth
approximate the denominator of the right-hand side of Eq. (\ref{stiffE}) \
by $(1-|\lambda |\rho _{1})^{3}r_{b}$. We also approximate the square-root
on the left-hand side of Eq. (\ref{stiffE}) by $(1-|\lambda |^{2}\rho
_{2}^{2})^{1/2}$. \ The stiff-limit is obtained by substituting orbit (\ref%
{multiscalesolution}) with $\rho _{k}=\sup (|R_{k}^{z}|,|R_{k}^{xy}|)$ into
Eq. (\ref{stiffE}), and taking the Fourier component of \ Eq. (\ref{stiffE})
along the PP frequency $w_{PP}\simeq $ $\pi q\Omega /\theta $. We also
multiply Eq. (\ref{stiffE}) by $r_{b}^{2}$, so that the order-of-magnitude
balancing posed by the Fourier-transformed version of Eq. (\ref{stiffE}) is%
\begin{equation}
\frac{m_{2}}{\mu \theta ^{2}}\rho _{2}\simeq \frac{(1-|\lambda |^{2}\rho
_{2}^{2})^{1/2}}{(1-|\lambda |\rho _{1})^{3}}\rho _{1}\exp (-\lambda ).
\label{STIFFLIM}
\end{equation}%
The equation of motion for the electron is obtained by interchanging indices
in Eq. (\ref{STIFFLIM}). A salient feature of Eq. (\ref{STIFFLIM}) is that $%
\lambda $ must be purely imaginary and multiple of $\pi $, $\ \lambda =\pi
qi $, so that the $\rho ^{\prime }s$ can be real in Eq. (\ref{STIFFLIM}),
i.e., the phase-shift $2\func{Im}(\lambda )$ must be a multiple of $2\pi $.
Considering the fraction on the right-hand-side of Eq. (\ref{STIFFLIM}) as a
given number, Eq. (\ref{STIFFLIM}) and the corresponding equation for the
electron are two linear homogeneous equations for $\rho _{1}$ and $\rho _{2}$%
. Equating the determinant of this homogeneous system to zero and
disregarding smaller terms yields 
\begin{equation}
\frac{\mu \theta ^{4}\exp (-2\lambda )}{M}\simeq \frac{(1-|\lambda |\rho
_{1})^{5/2}(1-|\lambda |\rho _{2})^{5/2}}{\sqrt{(1+|\lambda |\rho
_{1})(1+|\lambda |\rho _{2})}}.  \label{stiffdet}
\end{equation}%
Equation (\ref{stiffdet}) gives the main finite-amplitude correction to Eq. (%
\ref{Istar}), to which it reduces for $\rho _{1}=\rho _{2}=0$. The finite-$%
\rho $ corrections cancel the $\sigma ^{\prime }s$, which was the
justification for the approximation explained above Eq.(\ref{approximation}%
). When $\sup (\rho _{1},\rho _{2})$ approaches $\frac{1}{|\lambda |}$ from
below, the right-hand side of Eq. (\ref{stiffdet}) matches the $O(\theta
^{4})$ left-hand side, so that Eq. (\ref{stiffdet}) accepts a purely
harmonic solution with $\lambda =\pi qi$, as we wanted to demonstrate here.

After solving for the ping-pong oscillation, the dynamics at the slow
frequency is the next-to-leading of the multiscale solution. For the
electronic motion, Eq. (\ref{newton}) has a zero-order dissipative force
along the unperturbed orbit, as illustrated in Fig. 3. This offending force
has a component along the electronic velocity that is given by\ 
\begin{equation}
r_{b}^{2}F_{r}=r_{b}^{2}\frac{2}{3}\dddot{x}_{1}=-\frac{2e^{2}}{3c^{3}}%
\Omega ^{2}r_{b}^{2}\dot{x}_{1}=-\frac{2}{3}\theta ^{3},  \label{offending}
\end{equation}%
where we multiplied the force by $r_{b}^{2}$, as in Eqs. (\ref{LDEqsi1}) and
(\ref{dissi1}). The Lorentz-force of the proton along the\emph{\ unperturbed 
}circular orbit is almost normal to the electronic velocity and does not
contribute much for the dissipation. On the equation of the proton, the main
offending force at zero-order is the delayed interaction with the electron,
instead of the much smaller protonic self-interaction. Using Page's series
in the same way of Ref. \cite{PRL}, we find an offensive zero-order force
against the protonic velocity of the same magnitude of force (\ref{offending}%
). It turns out that it is impossible to cancel the zero-order force (\ref%
{offending}) with the linear terms of the variational equations, which is
shown by averaging the linearized equations over the orbital period, an
averaging that yields a $4\times 4$ linear system with no solutions. The
balancing of the offensive zero-order force is nevertheless possible by
using the resonance between ping-pong modes that is introduced at \emph{%
quadratic} order by the spin-radiation term, as discussed below.

Multiplying the offensive force (\ref{offending}) by the velocity along the
circular orbit yields an $O(\theta ^{4})$ dissipated power. Again, the
discussion of the dissipated power must start from the leading fast
dynamics, as follows. Taking the scalar product of Eq. (\ref{newton}) with\ $%
\mathbf{v}_{1}$and disregarding the contribution of the protonic fields for
the dissipated power yields 
\begin{equation}
m_{1}(\mathbf{v}_{1}\cdot \mathbf{a}_{1})=\frac{2}{3}\gamma _{1}\{(\mathbf{v}%
_{1}\cdot \mathbf{\dot{a}}_{1})+3\gamma _{1}^{2}(\mathbf{v}_{1}\cdot \mathbf{%
a}_{1})^{2}\}.  \label{power}
\end{equation}%
The bracket on the right-hand side of Eq. (\ref{power}) is multiplied by the
possibly large factor $\gamma _{1}$, so that it must vanish in the large-$|%
\mathbf{v}_{1}|$ limit, i.e.,

\begin{equation}
\ (\mathbf{v}_{1}\cdot \mathbf{a}_{1})^{2}\simeq -\frac{1}{3\gamma _{1}^{2}}(%
\mathbf{v}_{1}\cdot \mathbf{\dot{a}}_{1}).  \label{Otheta}
\end{equation}%
The offending force (\ref{offending}), multiplied by the slow guiding-center
velocity $\theta $ gives only an $O(\theta ^{4})$ contribution to the
right-hand side of Eq. (\ref{power}). The dominant contribution is given by
the PP oscillation, of frequency $\Omega |\lambda |/\theta $, and defined by
Eqs. (\ref{XYZPP}) and (\ref{multiscalesolution}). \ We henceforth replace
the $\mathbf{\dot{a}}_{1}$ on the right-hand-side of Eq. (\ref{Otheta}) by $%
-(\Omega ^{2}|\lambda |^{2}/\theta ^{2})\mathbf{v}_{1}$. Multiplying Eq. (%
\ref{Otheta}) \ by $r_{b}^{2}$ and using the fast component of orbit (\ref%
{multiscalesolution}) with $\rho _{1}\equiv \sup (|R_{1}^{z}|,|R_{1}^{xy}|)$
yields

\begin{equation}
|\lambda |^{2}\rho _{1}^{2}\cos ^{2}(\varkappa )\simeq \frac{|}{3}(1-\rho
_{1}^{2}|\lambda |^{2}),  \label{coefv1}
\end{equation}%
where $\varkappa $ is the angle between the velocity and the acceleration.
In the limit when $|\mathbf{v}_{1}|\rightarrow 1$, this angle $\varkappa $
must be close to ninety degrees, since $|\mathbf{v}_{1}|$ must be less than
one and $(\mathbf{v}_{1}\cdot \mathbf{a}_{1})=\frac{d}{dt}(\frac{|\mathbf{v}%
_{1}|^{2}}{2})$. Using Eq. (\ref{coefv1}) together with $|\cos (\varkappa
)|\leq 1$ yields a \emph{lower-bound} for $\rho _{1}$ 
\begin{equation}
\rho _{1}\geq \frac{1}{2|\lambda |}.  \label{relativeradius}
\end{equation}%
Next-to-leading in the multiscale solution is the guiding-center dynamics,
i.e., the equilibration of offending-force (\ref{offending}), a balance that
takes place \emph{after} the fast dynamics is established as a harmonic
oscillation with a radius given by Eq. (\ref{relativeradius}). In the
following we give a third derivation of resonance (\ref{resonancePPC}) by
averaging the equations of motion over the fast timescale. Due to the larger
protonic mass, we henceforth assume that the protonic field at the electron
averages essentially to the Coulombian field, a force that rotates at the
guiding-center frequency. On the other hand, in the limit $\rho _{1}|\lambda
|\rightarrow 1$, the electronic field on the proton, averaged over the fast
timescale, is significantly changed. Analogously to the derivation of Eq. (%
\ref{angularmomentum}), the electronic spin-radiation field, second term on
the right-hand side of Eq. (\ref{stiffE}), averaged over the fast timescale,
rotates at the guiding-center frequency $\Omega $ \emph{if} resonance (\ref%
{resonancePPC}) holds. This so averaged field adds to the rotating
electronic Coulombian field (not included in Eq. (\ref{stiffE})), so that
the averaged attractive force on the proton deviates from the diametral
direction. The proton repositions along the circular orbit, until its
velocity becomes \emph{perpendicular} to this perturbed centripetal force,
as illustrated in Fig. 3. After repositioning, the angular distance between
particles at the\emph{\ }same time is \emph{less} than $180$ degrees. The
self-consistent repositioning stops when the Coulombian field of the proton
along the electronic velocity acquires a component to balance the electronic
self-interaction (\ref{offending}). The perturbed equilibrium position along
the guiding-center orbit is illustrated in Fig. 3. The situation of Fig. 3
is possible only at resonance, when the averaged spin-radiation force acting
on the proton rotates at the guiding-center frequency. In Ref. \cite{PRL} we
also used a resonance between mutually orthogonal vibration modes. The
present work goes beyond our simple estimates of the helium dynamics of Ref. 
\cite{PRL}. Unlike the Coulombian orbits of helium\cite{PRL}, Coulombian
circular orbits of hydrogen radiate in dipole, and we have seen here that
the fast PP oscillations are essential to cancel this dipolar radiation, a
complex dynamics that demands a multiscale solution. The PP modes are lost
when the delay is expanded, as in Page's series of Ref.\cite{PRL}, a
nontrivial manifestation of stiffness.

For $\theta $ in the atomic magnitude, lower-bound (\ref{relativeradius}) is
a few percent of the interparticle distance, nevertheless the distance $\rho
_{1}r_{b}$ is \ already some $1000$ classical electronic radii, much larger
than the radius of the fat Lorentz electron\cite{Tomonaga}. To improve on
estimate (\ref{relativeradius}) one needs the relation between $\cos
(\varkappa )$ and $(1-\rho _{1}^{2}|\lambda |^{2})$, which is beyond the
present work. To solve Eq. (\ref{stiffdet}) it suffices that either $%
|R_{1}^{z}|$ or $|R_{1}^{xy}|$ approaches the upper limiting value $\frac{1}{%
|\lambda |},$ the other $R_{k}$ can take a much lower value, i.e., a
solution exists already in the limit when just the electron rotates near the
speed of light. As an estimate for the angular momentum of gyration, Eq. (%
\ref{angularmomentum}), we use $|R_{1}^{z}|$ and $|R_{1}^{xy}|$ given by Eq.
(\ref{relativeradius}), yielding 
\begin{equation}
\ |l_{xy}|=\frac{1}{4\pi \theta ^{2}q}.  \label{angularRxRy}
\end{equation}%
Calculating angular momentum (\ref{angularRxRy}) with the first line of
Table 1 gives $|l_{xy}|=213.5$. The angular momenta estimated by Eq. (\ref%
{angularRxRy}) depend on $q$, but are in the correct order of magnitude. The
electronic spin-angular-momentum of QED is independent of the orbital
quantum number and given by $|s|=\sqrt{3}\hbar /2=118$.

\section{Captions}

Fig. 1: Guiding-center circular orbit (dark lines) and particle trajectories
gyrating about the guiding-center (grey lines) for a three-frequency orbit 
\emph{near} resonance. Illustrated is also the beat of the electron's
gyration radius at \emph{about} the orbital frequency. Drawing not on scale;
the beat of the protonic gyration radius is not illustrated. Arbitrary units.

\bigskip

Fig. 2: Coulombian circular orbit with particles in diametral opposition at
the same time of the inertial frame (empty circles). Also indicated is the
retarded position of both particles (solid circles), and the angle travelled
during the light-cone time. Drawing not on scale; The circular orbit of the
proton and the retardation angle have an exaggerated magnitude for
illustrative purposes. Arbitrary units.

\bigskip

Fig. 3: Perturbed guiding-center orbit, with particles no longer in
diametral opposition at the same time. Positions at the same time are the
solid circles, the angular distance at the same time is $(\pi -\delta )$. At
the retarded position of the proton (empty circle), averaged force $F_{12}$
is perpendicular to the guiding-center velocity of the proton. At resonance,
averaged force $F_{12}$ rotates at the guiding-center frequency. Protonic
orbit and self-force have an exaggerated magnitude for illustrative
purposes. Arbitrary units.

\end{document}